\documentclass[a4paper, 12pt]{article}

\usepackage{graphicx}
\usepackage{color}
\usepackage{amsmath, amssymb, amsthm, ascmac, authblk, bm, booktabs, caption, comment, enumerate, latexsym, lscape, longtable, mathabx, mathrsfs, mathtools, multirow, natbib, newtxtext, psfrag, setspace, subcaption, thmtools, paralist, enumitem}
\usepackage{tabularx}
\usepackage[figuresright]{rotating}
\usepackage[stable]{footmisc}
\definecolor{mycolor}{cmyk}{0.85, 0.21, 0, 0.06}
\definecolor{applegreen}{rgb}{0.55, 0.71, 0.0}
\definecolor{blush}{rgb}{0.87, 0.36, 0.51}
\usepackage[colorlinks=true, linkcolor=mycolor, filecolor=mycolor, urlcolor=mycolor, citecolor=mycolor, backref = page, driverfallback=dvipdfmx]{hyperref}
\usepackage[margin = 2.3cm]{geometry}

\usepackage{xr}
\usepackage{clipboard}
\usepackage{ragged2e}
\usepackage{xurl}

\DeclareMathOperator*{\argmin}{argmin}

\DeclareMathOperator*{\obs}{\mathrm{obs}}
\DeclareMathOperator*{\ST}{\mathrm{ST}}
\DeclareMathOperator*{\KA}{\mathrm{KA}}

\DeclareMathOperator*{\pc}{\mathrm{c}}

\renewcommand{\hat}{\widehat}

\allowdisplaybreaks[1]

\theoremstyle{definition}

\declaretheorem[style=definition]{remark}

\numberwithin{equation}{section}

\definecolor{myred}{rgb}{0.76, 0.23, 0.13}

\makeatletter
\newcommand*{\rom}[1]{\expandafter\@slowromancap\romannumeral #1@}
\makeatother

\newcommand*\diff{\mathop{}\!\mathrm{d}}

\title{\textbf{Accessibility to Higher Education and Inter-regional Mobility} \vspace{.5cm}}

\author{
    Se-il Mun\thanks{Doshisha Business School, Doshisha University, Japan. }
    \hspace{1em}
    Shun Fujita\thanks{Kyoto Municipal Saikyo High School, Japan. }
    \hspace{1em}
    Lei Qin\thanks{Faculty of International Economic Studies, University of Niigata Prefecture, Japan. }
}

\date{\today}

\begin{document}

\onehalfspacing

\maketitle

\begin{abstract}
We develop an equilibrium model of the undergraduate higher education market that incorporates interregional mobility for enrollment.
An individual chooses a college to attend, seeking to maximize utility depending on the quality of education and the costs of attendance.
Costs include tuition fees, the effort required to pass the entrance exam, and moving costs from one's hometown to the region where the college is located.
Each college admits students whose entrance exam scores exceed a certain threshold, which we call the college's qualifying score.
In our model, the qualifying score of each college is determined endogenously as the shadow price of the constraint that the number of enrolled students must not exceed capacity.
We structurally estimate the model's parameters by simulating equilibrium using Japanese data.
Based on empirical results and counterfactual experiments, we evaluate the welfare costs of disparities in access to higher education due to uneven spatial distribution of colleges among regions.

\bigskip


\bigskip


\end{abstract}

\newpage

\section{Introduction}

\begin{figure}[tp]
    \begin{minipage}[b]{\textwidth}
        \centering
        \includegraphics[width=0.95\linewidth]{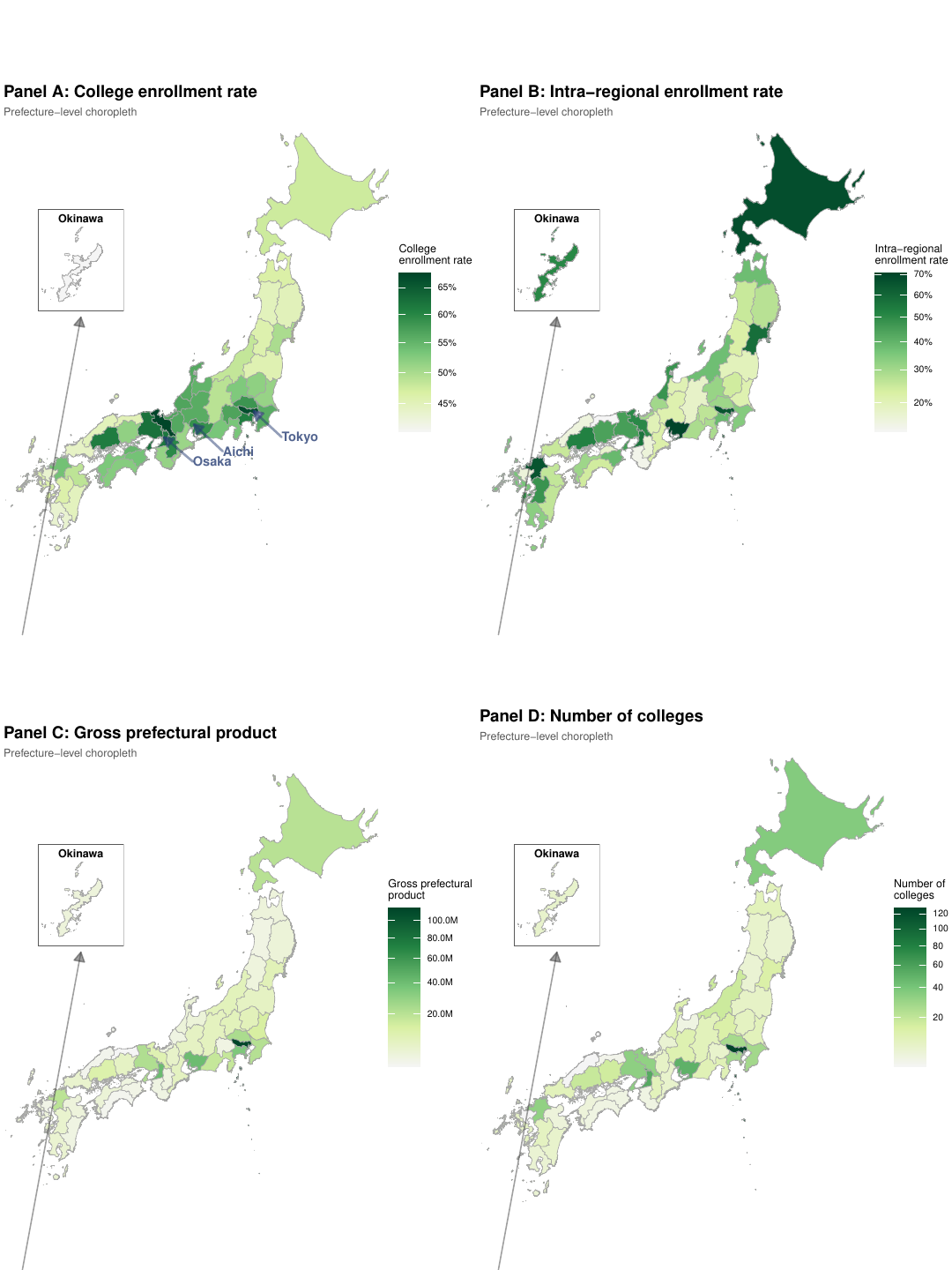}
        \captionsetup{width=.95\textwidth, font=footnotesize}
        \captionof{figure}{
        Prefecture-level comparison of college enrollment rate, intra-regional enrollment rate, Gross Prefecture Product (GPP), and the number of colleges in 2019.
        GPP data are from the Japanese Cabinet Office (\url{https://www.esri.cao.go.jp/jp/sna/data/data_list/kenmin/files/contents/main_2020.html} in Japanese).
        }
        \label{fig:prefecture_comparison}
    \end{minipage}
\end{figure}

There are substantial differences in the attributes of higher education enrollment across regions within a country. 
For instance, Panel A of Figure \ref{fig:prefecture_comparison} shows the variation in college enrollment rates\footnote{Defined as the number of college enrollments divided by the number of high school graduates. } across Japanese prefectures\footnote{Throughout this paper, we utilize the term ``college'' to refer to four-year undergraduate programs for simplicity.}. 
Referring to the regional income distribution in Panel C, we see that college enrollment rates are high in urban, developed prefectures and low in remote, peripheral ones\footnote{In the panel showing gross prefectural product, the prefectures with the higher incomes are Tokyo, Kanagawa, Osaka, and Aichi. Each of these prefectures has highly developed cities.}. 
Panel B depicts the intra-regional enrollment rate\footnote{Defined as the number of hometown enrollments divided by the total number of college enrollments. }, reflecting the propensity to attend college in one's home region. 
People in urbanized prefectures are more likely to choose colleges in their home regions, while those in remote prefectures are more likely to attend colleges outside their home regions\footnote{Regions are represented by prefectures. Japan has 47 prefectures. }.

Accordingly, the following question arises: 
If we assume that the distribution of intellectual ability among students is identical across regions, then why do such gaps occur between urban and remote areas of prefectures?

A plausible explanation centers on students' specific interests or the sought quality of higher education, influencing their choice of college.
Students unable to find a college that satisfies their criteria within a commutable distance may face the necessity of relocating, thereby incurring moving costs.
These costs encompass financial expenses and the psychological stress associated with adapting to a new cultural environment and unfamiliar local contexts.
Consequently, students from urban areas are more inclined to enroll in colleges within their home region, as urban regions have a greater number and broader range of colleges than remote areas.
This same accessibility advantage contributes to higher college enrollment rates in urbanized areas.

This disparity stems from the uneven geographic distribution of universities and colleges, as shown in Panel D of Figure \ref{fig:prefecture_comparison}.
Japan has 759 colleges in 2019, and as the most well-developed area in the country, the Tokyo Metropolitan Area (including Tokyo, Kanagawa, Saitama, and Chiba) has 27.8\% of them.
Another two main developed areas, the Kyoto-Osaka-Kobe Metropolitan Area and Nagoya Metropolitan Area account for 16.1\% and 6.7\%, respectively.
The remaining 49.4\% of colleges are located in the remaining 40 prefectures.
See \ref{ap:subsec:college} for further discussions about the uneven distribution of colleges in Japan.

This study aims to evaluate the welfare cost associated with disparities in higher education accessibility based on an equilibrium model of undergraduate higher education that incorporates inter-regional mobility for enrollment. Individuals select colleges based on a utility maximization framework, wherein the utility depends on the quality of education and the costs of attendance. These costs encompass tuition fees, the effort required to pass entrance exams, and costs associated with relocating from their hometown to the college's location. Colleges, on their part, admit students who achieve scores above a certain threshold in the entrance examination, referred to as the qualifying score. Our model endogenously determines each college's qualifying score as the shadow price for the enrollment capacity constraint.

We use data on inter-regional movement for higher education attendance in Japan to estimate the model's structural parameters and conduct a counterfactual experiment that removes the moving cost based on the estimated structural parameters.
This experiment demonstrates that eliminating moving costs could substantially reduce regional disparities in higher education accessibility, enabling students to choose colleges that match their abilities.

\textbf{Related literature.} This study relates to the literature on equilibrium models of the higher education market \citep*{epple2006, EPPLE2017, FU2014, cook2021competing, Fillmore2023}.
These models describe student choices, college admissions, and tuition fees; predict how students sort themselves among colleges based on income and ability; and establish a hierarchy of colleges by education quality.
They use such models to estimate structural parameters and conduct counterfactual analyses of changes in market design.
For example, \cite*{epple2006, EPPLE2017} examine the effect of removing price discrimination by colleges.
\cite{FU2014} examines the effects of tuition discounts and capacity expansions, as well as a setting wherein colleges have limited information regarding student ability.
\cite{Fillmore2023} studies price discrimination and public policy in the U.S. college market.

Most of this earlier equilibrium literature does not treat spatial accessibility as a central determinant of college choice\footnote{
\cite*{epple2006} aggregate private colleges into six groups and treat each group as one college. }.
Therefore, these works do not tackle the issue of regional disparity, wherein students from less accessible areas are required to relocate to access higher education opportunities.
The additional costs of inter-regional mobility can significantly impact college choice decisions.

Relatively recent studies have paid greater attention to the spatial distribution of colleges and geographic access to higher education \citep*{Gibbons2012, Newbold2015, cook2021competing, FU2022, Fabre2023, ishimaru2025}.
In England, \cite{Gibbons2012} demonstrate that geographic distance minimally impacts participation in higher education, but plays a key role in institution choice. In Canada, \cite{Newbold2015} examine the urban-remote gap in university attendance and report that local labor-market conditions, especially the local share of degree holders, help explain this gap alongside student, family, and access variables.
\cite{cook2021competing} investigates universities' competition in terms of prices, admission standards, and major program offerings.
In this work, the author models students' choice sets as being determined by their geographic locations.
\cite*{FU2022} introduce the distance between students' homes and their chosen colleges into the college choice model, exploring how home location influences expected utility across different income and ability levels.
\cite{Fabre2023} structurally studies how the geographic distribution of higher education contributes to spatial inequalities in France, demonstrating that minimizing mobility barriers improves welfare and educational attainment and strengthens the spatial concentration of human capital.
\cite{ishimaru2025} studies childhood locations and adult economic outcomes, highlighting the importance of spatial gaps in college and labor market opportunities.

Additionally, we relate to the literature on admission systems based on entrance examinations.
\cite{Hafalir2018} compare centralized and decentralized college admissions systems with entrance exams and study their equilibrium and welfare implications.
This perspective is particularly relevant for East Asian countries such as Japan, China, and South Korea, where entrance examinations play a major role in admissions.

Our model differs from previous studies in several ways.
First, we highlight inter-regional mobility's role within an equilibrium framework to evaluate the welfare costs of disparities in accessibility to higher education among regions.
Second, our model is tailored to examination-based higher education admissions systems, in which entrance examinations serve as the primary mechanism for allocating students among colleges.
Such systems are common in East Asian societies, including Japan, China, Taiwan, and South Korea\footnote{See \ref{ap:subsec:comparison} for discussions on the similarity between China's higher education system and that of Japan, and how our model is applicable to the case of China. }.
Compared with the United States, tuition plays a less prominent role in the allocation of students across public and private colleges in these systems.
Within this context, we introduce the concept of effort costs to achieve higher entrance exam scores.
Unlike previous models, such as \cite{EPPLE2017}, students are not sorted based on their innate abilities.
Students with lower abilities may gain admission to colleges with a higher quality of education by investing greater efforts than students with higher abilities.
Third, we introduce an empirical methodology that aligns closely with our theoretical framework, allowing for the estimation of structural parameters using aggregate data on student mobility across Japanese regions.

This paper proposes a microfounded structural model of the higher education market, but estimates it using aggregate data.
By formalizing the behavior of students and colleges, we derive a market-clearing condition and solve numerically for the equilibrium allocation of students across colleges.
Then, we estimate the structural parameters by matching model-implied aggregate enrollment flows to their empirical counterparts through a minimum-distance (M-estimation) approach.

This strategy differs from a method adopted in many quantitative spatial models, such as \cite*{ahlfeldt2015} and \cite{Redding2017}.
We discuss this point in Section \ref{subsec:estimation}.

Within the context of Japan's admission system, researchers in educational sociology have explored the factors influencing higher education attendance (cf. \citealp{Yamamoto1979}).
Through surveys and interviews, \cite{Hasegawa2016} highlights that the existence of various risks in modern society prevents students from going to college.
\cite{Yano2015} utilized data on 4,000 high school students and their parents to identify significant predictors of educational pathways, including parental education levels and household income.
\cite{Yonezawa2023} investigates the impact of economic and fiscal policies on Japan's higher education and discusses its future perspective under the current global challenges.

\cite{Muta1986} estimates a gravity model using time-series data and shows that the dispersion policy has rather enlarged a gap in the rate of university-bounded students.
\cite{Asia} uses a modified gravity model and analyzes the inter-prefectural movements related to entrance into colleges.
We use the same data of inter-prefectural movements as above, but develop a micro-economic model that enables welfare evaluation.

\textbf{Outline.} The remainder of the paper is structured as follows. Section \ref{sec:model} introduces the higher education market model. Section \ref{sec:application} outlines the data and methodology used for model estimation and evaluates the model's fit with the observed data. Section \ref{sec:counterfactual} discusses the welfare evaluation methodology and the impacts of the counterfactual experiment that removes moving costs. Section \ref{sec:conclusion} concludes the paper.

\section{The Model} \label{sec:model}

We consider a model containing $I$ regions, with each region $i \in \{1, 2, \dots, I\}$ having a number of students $L_i$ poised to graduate from high school.
Each of these students is confronted with the decision of whether to pursue higher education and, if so, which college to attend.
The system encompasses $N$ colleges, and each college $n \in \{1, 2, \dots, N\}$ is endowed with an exogenous capacity $K_n$ for enrolling new students.

Admission to any given college depends upon a student's performance in the entrance examination.
Specifically, a student must achieve a score equaling or exceeding the college's qualifying threshold.
The qualifying score is an equilibrium outcome that balances the demand for college seats with the supply of seats given by each college’s capacity.
Essentially, our model treats the qualifying score as the shadow price associated with the college-capacity constraint.
This approach ensures that students are allocated to colleges according to their ability to satisfy the academic requirements, as reflected by their examination scores, within the limitations imposed by available resources and geographic distribution.

\subsection{Individual decision-making} \label{subsec:decision_making}

There is a continuum of students who are heterogeneous with respect to their innate ability level,
denoted by $a$.
The distribution of ability is given by a density function,
$g(a)$,
which is identical across regions.
Following \cite{epple2006},
we assume that the decision to attend college is made by the student's household.
The utility function of a household of a student who attends a college $n$ is given by $u(c, \varphi_{n})$,
where $c$ is a num\'{e}raire consumption and $\varphi_{n}$ is the quality of education in college $n$.
We assume the utility is increasing in the consumption and the education quality:
\begin{equation*}
\frac{\partial u}{\partial c}>0,\quad \frac{\partial u}{\partial \varphi_{n}}>0.
\end{equation*}
The second inequality implies that attending a college with higher quality results in a higher return from education, and thereby contributes to a larger utility.
The quality of education in college $n$, $\varphi_{n}$, depends on the
quality of student body $e_{n}$, and educational resources $R_{n}$:
\begin{equation*}
\varphi_{n}=\varphi (e_{n},R_{n}).
\end{equation*}
We assume that the education quality $\varphi$ is increasing in the student quality $e_{n}$ and educational resources $R_{n}$:
\begin{equation*}
\frac{\partial \varphi }{\partial e_{n}}>0,\quad \frac{\partial \varphi }{\partial
R_{n}}>0.
\end{equation*}
The dependence of the education quality $\varphi_{n}$ on the student quality $e_{n}$ represents the \textit{peer effect} in the literature (cf. \citealp{sacerdote2011}).
A student would benefit from interacting with schoolmates or dormitory residents who have a strong intellectual background.
We assume that the minimum qualifying score for passing the entrance examination reflects the quality of the student body as a whole.
Specifically, we use the variable $e_n$ to represent both the minimum qualifying score and the quality of the students.
Educational resources include physical factors (such as monetary expenditure for education, and building facilities) and human resources (number and quality of faculties).
Thus, $R_{n}$ can be a composite of multiple variables.

The household's budget constraint is given by
\begin{equation}
y=c+\tau _{n}+s_{in}+h(a,e_{n}),
\label{eq:budget}
\end{equation}
where $y$ represents household income, $\tau_{n}$ the tuition, $s_{in}$ the cost of moving from the home region $i$ to the region where college $n$ is located, and $h(a,e_{n})$ the effort cost required to obtain the qualifying score $e_{n}$ for admission to college $n$.
We assume that the effort cost $h$ decreases with student ability $a$, exhibiting diminishing marginal returns, and increases with the qualifying score $e_n$, with increasing marginal cost:
\begin{gather*}
    \frac{\partial h}{\partial a} < 0,\quad \frac{\partial^2 h}{\partial a^2} > 0,\quad \frac{\partial h}{\partial e_{n}} > 0,\quad \frac{\partial^2 h}{\partial {e_n}^2} > 0.
\end{gather*}

In essence, our model posits that a student with higher innate ability may achieve a given qualifying score with less effort compared to a student with lower ability.
Conversely, obtaining a higher score necessitates increased effort.
This key distinction from the model presented by \cite{epple2006} is in the differentiation between a student's inherent ability and the score attained through effort\footnote{\cite{epple2006} suppose that peer student measure is represented by the mean ability level in the student body. Their model employs SAT scores as the measure of student ability. }.
This nuance mirrors the reality faced by students within the college entrance examination framework of Japan and other East Asian societies, wherein extensive preparation typically begins early.
Students frequently enroll in preparatory schools and dedicate considerable hours to studying, occasionally at the cost of leisure activities.
Consequently, a student with a lower innate ability level may surpass another student with a higher innate ability by investing significant effort\footnote{This formulation implies that students choose a level of effort to obtain a score just slightly above the qualifying score for college. In fact, the distribution of admission scores is concentrated around the qualifying score. }.
This phenomenon reveals the importance of effort and preparation in the context of East Asia's educational attainment, highlighting a critical aspect of student performance and college admissions.

The moving costs $s_{in}$ encompass more than mere financial costs associated with relocation. Moreover, it includes the psychological burden arising from anxiety regarding adapting to a different culture, navigating an unfamiliar environment, and the challenge of being away from a familiar support network. We assume that the moving cost is positively correlated with the distance between a student's home region and the college’s location:
\begin{equation*}
s_{in}=\sum_{j=1}^{I}\alpha _{nj}f(d_{ij}),
\end{equation*}
where $d_{ij}$ is the distance between regions $i$ and $j$,
and $f(d_{ij})$ is a monotonic increasing function of distance. $\alpha _{nj}$ is a dummy variable indicating that college $n$ locates in
region $j$,
such that
\begin{align*}
\alpha _{nj}\coloneqq
\begin{cases}
1 & (\text{if college }n\text{ locates in region }j) \\
0 & (\text{otherwise})
\end{cases}.
\end{align*}

Incorporating the budget constraint into the utility function to cancel $c$, we have the indirect utility
\begin{equation}
V_{in}(a)
=u\left( y - \tau_n - s_{in}-h(a,e_{n}), \varphi (e_{n}, R_{n}) \right).
\label{eq:indirect_utility}
\end{equation}
The household has the outside option of not attending college, which is defined as $n = 0$.
The utility level for this outside option is represented by a constant, such that $V_{i0}(a) = u_{0}$.

Each household chooses one alternative $n \in \{0, 1, 2, \dots, N\}$ to maximize its utility.
Further, we assume that each household in region $i$ has its own preference for college $n$, which is described by adding a stochastic term,
$\varepsilon _{in}$.
The condition for utility maximization is written as follows.
\begin{equation*}
V_{in} (a) + \varepsilon_{in} > V_{im}(a) + \varepsilon_{im}, \quad \forall m \neq n,
\end{equation*}
Assuming that $\varepsilon _{in} \sim \text{Gumbel}(0, 1 / \lambda)$ is i.i.d. for all $i$ and $n$, we obtain the logit model for the probability that a student with ability $a$ from region $i$ enrolls the college $n$, as follows
\begin{equation}
P_{in}(a)=\frac{\exp {\left(\lambda V_{in}(a)\right)}}{\sum_{m=0}^{N}\exp {\left(\lambda V_{im}(a)\right)}}.
\label{eq:probability}
\end{equation}
The probability that a student chooses the outside option is
\begin{equation}
P_{i0}(a) = \frac{\exp(\lambda u_0)}{\sum_{m=0}^{N}\exp (\lambda V_{im}(a))}.
\label{eq:probability_n0}
\end{equation}

\subsection{Equilibrium allocation of students among colleges}

Within the context of the college market, wherein students demand and colleges supply educational services, achieving market equilibrium involves appropriately allocating students across all colleges such that no college exceeds its enrollment capacity.
A distinctive feature of the Japanese higher education system is that tuition fees are not subject to fluctuation based on the demand-supply dynamics; notably, all national universities charge the same tuition fee.
Consequently, the price of educational services in colleges is not market-driven.

The essential mechanism in this market is the entrance examination score instead of the price. Colleges organize these exams for prospective students and select them based on their scores. As students naturally gravitate toward higher-quality colleges, such institutions typically receive more applications than they can accommodate, fostering a competitive selection process. Additionally, the geographic location of colleges is important; colleges located in metropolitan areas tend to attract more applicants, necessitating higher qualifying scores for admission.

Thus, similar to how prices work in conventional goods and services markets, entrance examination scores serve to balance demand with supply in the college market. They effectively allocate students among colleges, ensuring that enrollment aligns with college enrollment capacities while also reflecting the applicant pool’s preferences and abilities.

The number of students from region $i$ who choose to enroll in college $n$ is written as
\begin{equation*}
F_{in}=L_{i}\int_{\underline{a}}^{\overline{a}}P_{in}(a)g(a) \diff a.
\end{equation*}
Note that the domain of the density function $g$ is defined as $[\underline{a}, \overline{a}]\subset \mathbb{R}$, where $\underline{a}$ and $\overline{a}$ represent the lower and upper bounds, respectively.
By aggregating $F_{in}$ for all $i \in \{1, 2, \dots, I\}$, we have the demand for college $n$, denoted by $E_{n}$:
\begin{equation}
E_{n} = \sum_{i=1}^{I}F_{in} = \sum_{i=1}^{I}L_{i}\int_{\underline{a}}^{\overline{a}}P_{in}(a)g(a) \diff a.
\label{eq:enrollment}
\end{equation}
Notice that $E_{n}$ represents the number of students who choose to attend college $n$.

In our model, the capacity $K_n$ of each college, or the maximum number of students that can be enrolled, is given exogenously, mirroring the situation in Japan where college capacities are regulated by governmental oversight.
A specific government council evaluates and approves applications from colleges to modify their capacities.
To ensure that the system operates within these constraints, the following inequality should hold for all $n$:
\begin{equation}
    K_{n} \geq E_{n}.
    \label{eq:capacity}
\end{equation}
At equilibrium, the qualifying-score vector is denoted by
$\mathbf{e} \coloneqq (e_{1}, e_{2}, \dots, e_{N})^\top$. 
Qualifying scores are adjusted to satisfy the condition specified in \eqref{eq:capacity}.
If demand surpasses capacity at certain colleges, the qualifying scores at those colleges are increased accordingly, and adjustments may also be made to the scores at other colleges.

To close the equilibrium system, we assume that each college fills its enrollment capacity.
To realize this motivation, college $n$ sets its qualifying score to precisely match the exam score of the student who ranks at the ($K_{n}$)th position among all exam-takers.
Following this motivation, each college makes the maximum utilization of the higher education seats to which it has been assigned, such that
\begin{equation}
    K_{n} = E_{n}
    \label{eq:market_clearing}
\end{equation}
for all $n$.
We call \eqref{eq:market_clearing} the \textit{theoretical market-clearing condition}, as it yields the equivalence between the higher education supply ($K_n$) and demand ($E_n$).

Noteworthily, since the tuition and the capacity of colleges are considered exogenous in this paper, this assumption is equivalent to the behavior of tuition revenue maximization of colleges, because the tuition revenue is maximized if and only if all seats are fulfilled.

\begin{remark} \label{rm:motivation}
    Profit maximization is not the only consideration of college behaviors in the field of education economics.
    Common alternative assumptions of the colleges' goals include research quality maximization, students' welfare maximization, and financial sustainability.
    However, the assumption that colleges fill their capacities is crucial for determining equilibrium qualifying scores in our model.
    If colleges are allowed to raise their qualifying scores while leaving seats vacant, the capacity inequalities in \eqref{eq:capacity} alone do not pin down a unique qualifying-score vector $\mathbf{e}$.
\end{remark}

Our model does not explicitly describe the process by which colleges select students among applicants in order of their scores.
In reality, each student applies to multiple colleges by taking exams administered by these colleges and is accepted by some and rejected by others.
Instead of formulating the above process,
our model collapses these stages into a simple scenario, wherein students choose a level of effort to obtain the qualifying score $e_{n}$ of college $n$ and are actually admitted.

Note that $e_{n}$ itself contributes to the quality of colleges and raises the utility of students.
Colleges with higher levels of $e_{n}$ attract more students, resulting in more competitive admissions.
This increased competition further raises $e_{n}$ at those colleges, creating a positive feedback resembling the urban agglomeration.
Students face a trade-off in the sense that they benefit from the peer effects at colleges with higher $e_{n}$ but must incur greater effort costs to achieve a score at least equal to $e_n$ on the entrance exam.

\section{Application to Higher Education System in Japan} \label{sec:application}

This section introduces the details of the data of this research, identification, and estimation results. We also discuss the implications of the equilibrium allocation.

\subsection{Data} \label{subsec:data}

In the empirical part of this paper, regions are represented by prefectures of Japan ($I = 47$).

\textbf{Prefectural variables.}
The data of the inter-regional movement of students from prefecture $i$ to $j$, $M^{\obs}_{ij}$ is from `The number of enrollees by prefecture where senior high schools from which they graduated are located' in the Ministry of Education, Culture, Sports, Science and Technology (\url{https://www.mext.go.jp/b_menu/toukei/chousa01/kihon/1267995.htm} in Japanese).
The total number of people in prefecture $i$ who choose to attend a college is calculated by $l_i = \sum^{I}_{j = 1} M^{\obs}_{ij}$.
The population of high school graduates of each prefecture, ${L}_{i}$ is calculated reversely by $l_i$ and data on the college enrollment rate of each prefecture provided by the Portal Site of Official Statistics of Japan (\url{https://www.e-stat.go.jp/} in Japanese).
The distance between prefectures, $d_{ij}$ is from the data in the Geospatial Information Authority of Japan (\url{https://www.gsi.go.jp/gis.html} in Japanese).

The data on income is available at the Portal Site of Official Statistics of Japan (\url{https://www.e-stat.go.jp/} in Japanese).
In this paper, we use the disposable income per capita (OECD standard) of households with two or more people in year 2019.

\textbf{College groups.}
There are 786 colleges and 2,577 faculties as of 2020. For computational convenience, we aggregate these colleges by prefecture and college group of similar quality.

The rules of aggregation are as follows: First, for each prefecture, colleges are divided into public ones and private ones. In this procedure, `national universities' (universities that are established by the Japanese government) are classified as public colleges.

Second, for each prefecture where a \textit{former imperial university} is located\footnote{Hokkaido, Tohoku, Tokyo, Nagoya, Kyoto, Osaka, and Kyushu.}, we build two groups of public colleges.
The first public group contains only the oldest seven national universities, and the second public group involves the remaining public colleges in this prefecture.
For example, ``Tokyo Pub1'' refers to the University of Tokyo (one of the oldest seven national universities), and `Tokyo Pub2' contains all public colleges in Tokyo except for the University of Tokyo. For other prefectures, the public colleges are aggregated as one public group.

Furthermore, we set up three groups of private colleges in each of Tokyo, Kanagawa, Kyoto, and Osaka.
This is because these prefectures have a large number of private colleges, and their quality varies extensively.
Specifically, we divide college groups in these four prefectures according to the levels of qualifying scores\footnote{ The thresholds used to divide the groups are explained later. }.
For other prefectures, private colleges are aggregated into one group.
Owing to group division, colleges in Japan are aggregated into $N = 107$ groups.\footnote{
Some colleges have multiple faculties across prefectures.
Consequently, nearly 11\% of students attend campuses in the prefectures different from the ones with the headquarters of the colleges.
To circumvent potential statistical skews, the faculties are aggregated based on where their offices or their campuses for 1st-grade students are located. }

\textbf{College-related variables.}
Tuition data is from official announcements of the annual tuition of colleges.
We use the average tuition in faculties that charge multiple tuition prices for different majors and/or curricula.
The variable $\tau_n$ is the median tuition of aggregated faculties.
The median instead of the mean is used to alleviate the influence of high tuition fees by,
for example,
medical and pharmaceutical faculties.

The educational resource is a synthesized measure using the amount of Grants-in-Aid for Scientific Research (KAKENHI) and the student-teacher ratio. KAKENHI demonstrates the strength of faculty members' research. We assume that faculty members who have made significant research contributions could provide a higher quality education.
The KAKENHI data of fiscal year 2019 is from the Japan Society for the Promotion of Science (\url{https://www.jsps.go.jp/english/e-grants/rrda.html}).
For colleges that do not receive KAKENHI, their measure of revenue is set as JPY 15 million.
The student-teacher ratio is defined as
\begin{gather*}
    \text{student-teacher ratio} \coloneqq \frac{\text{number of students}}{\text{number of teachers}}.
\end{gather*}
A smaller student-teacher ratio indicates a smaller class size and hence a better education quality.
Data on the numbers of students and teachers are available on colleges’ websites.
The educational resource is calculated as:
\begin{equation}
    \begin{split}
    \text{educational resource} \coloneqq{}& \alpha \cdot \left( 1 - \frac{\ST_n - \min_{n} \ST_n}{\max_n \ST_n - \min_{n} \ST_n} \right) \\
    &{}+ (1 - \alpha) \cdot \frac{\KA_n - \min_n \KA_n}{\max_{n} \KA_n - \min_n \KA_n},
    \end{split}
    \label{eq:er}
\end{equation}
where $\ST_n$ and $\KA_n$ represent the college $n$'s student-teacher ratio and KAKENHI, respectively.
The synthesized educational resource ranges from 0 to 1,
with higher values indicating a better educational environment.
The weight $\alpha$ balances the importance of class size and research funding; in practice, we choose $\alpha = 0.10$ in terms of the comparison of education resource ranking and observed qualifying score ranking.
In \ref{ap:subsec:alpha}, we show exactly how we choose the value of $\alpha$ among all candidate numbers.
Each college’s capacity is derived from the official school guide of each college.
The 2021 data are utilized, and for the missing data, the 2020 data are used.
The values of the variable $K_n$ are the sum of all capacity values for colleges within the same college group discussed above.

Each college prepares its own entrance examination, and the scores are not comparable between different colleges.
However, the Japanese education industry has developed an index, \textit{`Hensachi'}, that provides the score level of each college in normalized units.
This normalized index has been widely used to rank each university by the score level required to pass the entrance examination.
See \ref{ap:subsec:entrance} for further discussions about Hensachi.
We use this index for the qualifying score $\mathbf{e}^{\obs} \coloneqq (e^{\obs}_{1}, e^{\obs}_{2}, \dots, e^{\obs}_{N})$, based on a survey conducted by Sundai College-Preparatory School and Benesse Corporation.
The private-college groups in Tokyo, Kanagawa, Kyoto, and Osaka are divided using the following thresholds: first-tier ($e^{\obs} \geq 60$), second-tier ($45 \leq e^{\obs} < 60$), and third-tier ($e^{\obs} < 45$).
The two companies together gathered the results from applicants and forecasted the trend of the college entrance examination.
This forecast captures the trends adequately in the sense that over 78\% of applicants participated in the project.
Some values are omitted in this project, and such omitted values are substituted by the investigation results of Kawaijuku College-Preparatory School. Data of $\{\tau_n, K_n, e^{\obs}_{n}\}$ are collected from \cite{obunsha}.

Table \ref{tab:describe} summarizes the definition and descriptive statistics of the collected data.
\begin{table}[t] 
    \begin{center}
        \caption{Descriptive Statistics of Variables}
        \label{tab:describe}
        \small
            \begin{tabular}{@{}lrrrrrr@{}}
            \toprule
            Variable & Min & Max & Mean & Std. Dev. & Units \\ 
            \midrule
            $y_i$ & 3.82 & 6.13 & 5.02 & 0.46 & 1M JPY \\
            $a$ & 50 & 150 & 100 & 15 & - \\
            $\tau_n$ & 0.54 & 1.44 & 0.69 & 0.19 & 1M JPY \\
            $R_n$ & 0.01 & 0.78 & 0.08 & 0.11 & - \\
            $K_n$ & 80 & 70978 & 5655.80 & 9766.56 & persons \\
            $e^{\obs}_{n}$ & 34.00 & 68.97 & 48.30 & 8.39 & - \\
            $L_i$ & 5125.84 & 119017.4 & 23686.36 & 23623.65 & persons \\
            $l_i$ & 2220 & 77516 & 13003.47 & 14843.21 & persons \\
            $d_{ij}$ & 0 & 22.44 & 5.09 & 3.59 & 100 km \\
            $M^{\obs}_{ij}$ & 0 & 51021 & 276.67 & 1767.54 & persons \\
        \bottomrule 
        \end{tabular}

        \begin{flushleft}
        Note: 
        $y_i$: income; 
        $a$: ability; 
        $\tau_n$: tuition; 
        $R_n$: education resource; 
        $K_n$: capacity; 
        $e^{\obs}_{n}$: qualifying score; 
        $L_i$: high school graduates; 
        $l_i$: students enrolled in colleges; 
        $d_{ij}$: distance; 
        $M^{\obs}_{ij}$: inter-regional movement. 
        Ability $a$ comes from assumption of IQ distribution, not from observed data. 
        $L_i$ is calculated reversely from the observed $l_i$ and the prefectural college enrollment rate. 
        \end{flushleft}
        
    \end{center}
\end{table}

\subsection{Functional Form Specification} \label{subsec:functional}

In this paper, we use the term Intelligence Quotient (IQ) to represent the ability level $a$.
Thus, it follows a truncated normal distribution with a mean of 100, a standard deviation of 15, and is bounded between 50 and 150: $a \sim \text{TN}(100, 15, 50, 150)$.

To calculate the equilibrium numerically,
we need to specify the functional form of the utility function of a household.
Specifically,
we assume a quasi-linear form of the utility function as follows:
\begin{gather}
    V_{in}(a) = y_i - \tau_n - s_{in} - h(a, e_n) + \varphi(e_n, R_n), \label{eq:quasi_ufunc} \\
    s_{in} = \theta_1 \sum_{j = 1}^{I} \alpha_{nj} d_{ij}, \quad \varphi(e_n, R_n) = \theta_3 \log(e_n) + \theta_4 \log(1 + R_n), \notag
\end{gather}
and
\begin{align}
    h(a, e_n) = \underline{I} + \theta_2 \cdot \frac{1}{(a - \underline{a})} \log \left( \frac{50}{\bar{e} - e_n} \right).
    \label{eq:education_investment}
\end{align}

We adopt a quasi-linear utility specification for two reasons.
First, it delivers a transparent welfare metric, which is particularly useful for the counterfactual analysis in Section \ref{sec:counterfactual}, because changes in utility can be interpreted in monetary terms without requiring additional normalization.
Second, it helps to maintain identification under the data available in this study.
In our setting, key sources of individual heterogeneity, most notably household income and student ability, are not observed at the individual level.
Instead, income is observed only at an aggregate regional level, while ability is modeled parametrically.
Under these data limitations, a more flexible utility specification would introduce multiple coefficients on latent or weakly observed individual characteristics, rendering it challenging to separately identify their marginal effects from aggregate enrollment flows alone.

Generally, when the data contain limited cross-sectional variation in the underlying individual-level state variables, richer nonlinear specifications can generate observationally similar choice patterns for different parameter combinations.
This weakens empirical discipline and makes structural interpretation less reliable.
The quasi-linear specification mitigates this problem by reducing the dimensionality of the preference system while preserving the key economic trade-offs among tuition, mobility cost, effort cost, and college quality.
Therefore, owing to the objectives of the paper and the level of aggregation in the data, the quasi-linear specification provides a tractable, interpretable, and empirically identifiable framework for the analysis.

The structural form of \eqref{eq:education_investment} satisfies the assumptions made in Subsection \ref{subsec:decision_making}.
It is specified such that the range of qualifying score distributions obtained by the model is consistent with that of the observed distributions, that is, between 30 and 80 (see \ref{ap:subsec:entrance} for details).
In the numerical simulation, we set $\bar{e}=80$ and let $\underline{I} = 450$ (thousand JPY) as one-year pecuniary cost of a public high school, including tuition fee, school education expenses, and extracurricular activity expenses.

\subsection{Estimation} \label{subsec:estimation}

The structural parameters are $\boldsymbol{\theta} = (\theta_1, \theta_2, \theta_3, \theta_4, u_0, \lambda)$.
Our model yields an equilibrium inter-regional flow as $M_{ij}(\boldsymbol{\theta})$ by the following equation:
\begin{equation*}
    M_{ij}(\boldsymbol{\theta}) \coloneqq \sum_{n=1}^N \alpha_{nj} F_{in}.
\end{equation*}
Notice that $F_{in}$ is the number of students graduating from high schools in prefecture $i$ attending college $n$ by model prediction. The structural parameters are estimated by minimizing the objective function, which is defined as the average sum of squared discrepancy of inter-regional enrollment of students between model prediction and observed data:
\begin{align}
    \hat{\boldsymbol{\theta}} \coloneqq \argmin_{\boldsymbol{\theta} \in \boldsymbol{\Theta}} \frac{1}{I^2} \sum_{i=1}^I \sum_{j=1}^I \rho_{ij} (\boldsymbol{\theta})
    \quad \text{with} \quad
    \rho_{ij} (\boldsymbol{\theta}) \coloneqq \left(M_{ij}(\boldsymbol{\theta}) - M^{\obs}_{ij} \right)^2. \label{eq:objective}
\end{align}
Notice that $\hat{\boldsymbol{\theta}}$ is an M estimator of $\boldsymbol{\theta}$.

\begin{remark} \label{rm:tradition}
    The traditional estimation strategy is to maximize the likelihood related to $P_{in}(a)$ (cf. Chapter 2 of \citealp{small2007}).
    Unfortunately, the individual-level data of $P_{in}(a)$ is not available. The estimation strategy of \eqref{eq:objective} follows the minimum distance or M estimation methodology related to the aggregated data of inter-regional movement $M_{ij}$.
    This method leverages the available data on the aggregate inter-regional enrollment of students, thus allowing for the estimation of structural parameters in the absence of individual-level data.
\end{remark}

A useful comparison is provided by inversion-based estimation in quantitative spatial economics. \cite{ahlfeldt2015} also work with aggregate spatial data; however, their empirical setting provides a particular set of observed equilibrium objects---including floor prices, workplace employment, residence employment, land area, and bilateral travel times---that map directly into the equilibrium conditions of their model.
Owing to these observables and a parameter vector, their model admits a one-to-one inversion from observed equilibrium outcomes to adjusted production and residential fundamentals, which can be used as structural residuals in GMM estimation.

By contrast, the model-data mapping in our setting does not support the same inversion strategy.
The main observed outcomes are aggregate inter-regional student flows, college capacities, entrance exam scores, and college or regional attributes.
These variables are informative about student sorting and equilibrium admissions outcomes, but they do not provide the full set of equilibrium objects needed to recover unobserved fundamentals through an analogous closed-form inversion.
Specifically, for each candidate parameter vector, the equilibrium qualifying scores must be solved forward from the college market-clearing conditions before model-implied enrollment flows can be compared with the data.
For this reason, we employ an M estimator with a nested numerical equilibrium solver: For each candidate parameter vector, we first solve for the equilibrium qualifying scores and then compare the resulting model-implied enrollment flows with the observed flows.

The advantage of using aggregate data is that such data are generally easier to obtain and, in our setting, provide sufficient information to discipline the mapping between observed inter-regional student flows, college capacities, qualifying scores, and model-implied equilibrium outcomes.
This makes aggregate data useful for quantifying the spatial distribution of equilibrium access to higher education and for conducting welfare analysis and policy counterfactuals.
The limitation is that, relative to studies using individual-level data (cf. \citealp{epple2006, FU2014}), aggregate data contain less information for identifying rich forms of individual heterogeneity.
Consequently, although the model is structural, some estimated parameters may still admit a reduced-form interpretation.

\begin{remark} \label{rm:m_etm}
    Noteworthily, the equilibrium qualifying score $\mathbf{e}^*$ is determined by both the structural parameters $\boldsymbol{\theta}$ and the initial values $\mathbf{e}^{(0)}$, which will be discussed later in \ref{step:equi:initial} of \ref{alg:inner}.
    As the model's prediction of inter-regional enrollments, $M_{ij} (\boldsymbol{\theta})$, also depends on $\mathbf{e}^*$, a more precise expression of $\rho_{ij} (\boldsymbol{\theta})$ should explicitly incorporate the dependency on $\mathbf{e}^*$, as follows:
    \begin{align}
        \rho_{ij} (\boldsymbol{\theta}) = \left(M_{ij}(\boldsymbol{\theta}, \mathbf{e}^*(\boldsymbol{\theta})) -  M^{\obs}_{ij} \right)^2.
        \label{eq:objective_rgr}
    \end{align}
    This rigorous expression is still categorized as M estimator. We learn from \eqref{eq:objective_rgr} that the structural parameters $\boldsymbol{\theta}$ exert both direct and indirect effects on inter-regional enrollments $M_{ij}$, the latter through their influence on the equilibrium qualifying score $\mathbf{e}^*$.
\end{remark}

The estimation follows an inner loop-outer loop approach.
The inner loop finds an endogenous qualifying score $\mathbf{e}^*$ for a given parameter vector, whereas the outer loop searches the parameter vector to minimize the objective function \eqref{eq:objective}.
The inner loop is described as follows:
\begin{enumerate}[leftmargin=\parindent,
    align=left,
    labelwidth=\parindent,
    label=\textbf{Inner Loop (Equilibrium Algorithm)},
    ref=\textbf{Inner Loop}]
    \item \label{alg:inner} \text{}
    \begin{enumerate}[label=\textbf{step A\arabic*}]
        \item \label{step:equi:initial} Choose the initial qualifying-score vector, denoted as
        $\mathbf{e}^{(0)} \coloneqq (e_{1}^{(0)},e_{2}^{(0)},\dots ,e_{N}^{(0)})^\top$.
        Set the iteration counter to $k = 1$.
        \item \label{step:equi:update} Calculate the enrollment number of colleges $n$ under $\mathbf{e}^{(k-1)}$, $E_n(\mathbf{e}^{(k-1)})$, using \eqref{eq:indirect_utility}-\eqref{eq:enrollment}, and update the value of $\mathbf{e}^{(k)}$ by the \textit{iteration equation}
        \begin{align}
            {e}_n^{(k)}={e}_n^{(k - 1)}+\zeta \left[ E_{n}(\mathbf{e}^{(k - 1)})-K_{n}\right]
        \label{eq:update}
        \end{align}
        for all $n$, where $k$ is the iteration counter.
        \item \label{step:equi:check} Repeat \ref{step:equi:update} until the \textit{numerical market-clearing condition}
        \begin{align}
                E_{n}(\mathbf{e}^{(k)}) \leq (1 + \varepsilon) \cdot K_{n}
            \label{eq:condition}
        \end{align}
        is satisfied for all $n$, or until $k = k_{\max}$, where $k_{\max}$ is the preset maximum number of iterations.
    \end{enumerate}
\end{enumerate}

Notice that the numerical market-clearing condition \eqref{eq:condition} is weaker than the theoretical one \eqref{eq:market_clearing} for computational reasons.
In reality, colleges frequently admit students more than their capacities, as admitted students may enroll in other colleges.
The Japanese government accepts the over-capacity rate of colleges below a specific threshold, and hence the setting of \eqref{eq:condition} is reasonable.
In the numerical simulation, we set $\varepsilon = 0.05$, $k_{\max} = 5,000$, and $\zeta = 1.0 \times 10^{-5}$.
For colleges such that $K_n \leq 600$, the RHS of \eqref{eq:condition} are replaced by $(1 + \varepsilon) K_n + 100$ to ease computational difficulties.
If \eqref{eq:condition} is satisfied, we judge that the numerical equilibrium is achieved and let $\mathbf{e}^* \coloneqq \mathbf{e}^{(k)}$.

Introducing peer effects into our model leads to the presence of multiple equilibria owing to externalities.
It implies that several values of $\mathbf{e}$ could satisfy the theoretical market-clearing condition \eqref{eq:market_clearing}.
To address this issue, our strategy is to focus on the equilibrium oriented from the observed qualifying score $\mathbf{e}^{\obs}$:
That is, fix $\mathbf{e}^{(0)} \equiv \mathbf{e}^{\obs}$ in \ref{step:equi:initial} in \ref{alg:inner}.
This kind of strategy dealing with multiple equilibria is common in the related literature, for example, \cite{EPPLE2017} focus on a “hierarchical adherence” equilibrium where private college hierarchy follows the endowment hierarchy.

This strategy's benefits are as follows:
(\romannumeral1) Starting the iteration from observed data means the model's convergence will likely reflect an equilibrium that is empirically relevant or plausible under current or observed market conditions.
(\romannumeral2) In markets with multiple equilibria, the equilibrium that is closest to the observed market condition can be thought of as the `natural' outcome of the price adjustment process.
(\romannumeral3) This approach is computationally efficient for analyzing equilibrium close to observed data, facilitating easier comparison between model predictions and real-world outcomes.

Nevertheless, it is worth highlighting this strategy's limitations and assumptions.
Specifically, it assumes that the observed data is sufficiently close to equilibrium or that the market dynamics modeled by the iteration equation adequately capture how the actual market adjusts.
Moreover, it is essential to consider whether this approach might bias the analysis towards equilibria that are more likely given current observations but potentially overlook other equilibria that could emerge under different conditions.

\begin{remark} \label{rm:multiequilibria}
    Another approach to addressing the problem of multiple equilibria is to allocate the set of all possible multiple equilibria and conduct a comprehensive analysis based on this set. For example, one can select a tolerance or criterion based on economic theory or statistical evidence. Then she can determine the confidence interval for structural parameters by including only those multiple equilibria that meet the predefined tolerance. However, it is computationally infeasible to obtain all possible multiple equilibria in this study.
\end{remark}

In the outer loop, we calculate the value of the objective function \eqref{eq:objective} by executing the \ref{alg:inner} under given values of $\boldsymbol{\theta} = (\theta_1, \theta_2, \theta_3, \theta_4, u_0)$ and $\lambda$.
Then, we minimize the objective function by adjusting $\boldsymbol{\theta} = (\theta_1, \theta_2, \theta_3, \theta_4, u_0)$ and $\lambda$ by some optimization rules defined by \eqref{eq:objective}.
This procedure is technically described as follows:
\begin{enumerate}[leftmargin=\parindent,
    align=left,
    labelwidth=\parindent,
    label=\textbf{Outer Loop (Parameter Search Algorithm)},
    ref=\textbf{Outer Loop}]
    \item \label{alg:outer} \text{}
    \begin{enumerate}[label=\textbf{step B\arabic*}]
        \item \label{step:opti:initiate} Fix an initiating point of $\lambda$ and $\boldsymbol{\theta} = (\theta_1, \theta_2, \theta_3, \theta_4, u_0) \in \boldsymbol{\Theta}$, where $\boldsymbol{\Theta}$ is a preset parameter space.
        Calculate $\rho_{ij}(\boldsymbol{\theta})$ at this initiating point via the \ref{alg:inner}.
        \item \label{step:opti:grid} For each fixed value of $\lambda$, evaluate the objective function at $B$ candidate parameter vectors $\boldsymbol{\theta} \in \boldsymbol{\Theta}$, where $B$ denotes the number of candidate combinations in the coarse grid ($B = 5 \times 5 \times 3 \times 5 \times 4 = 1{,}500$ in our implementation).
        Let $\check{\boldsymbol{\theta}}$ denote the candidate vector with the smallest objective-function value.
        \item \label{step:opti:optimization} Use derivative-based optimizing algorithm of statistical software to find $\hat{\boldsymbol{\theta}}$ that minimizes the objective function starting at $\check{\boldsymbol{\theta}}$, within a convex parameter space $\check{\boldsymbol{\Theta}}$ that contains $\check{\boldsymbol{\theta}}$.
    \end{enumerate}
\end{enumerate}

In \ref{step:opti:initiate} and \ref{step:opti:grid}, we restrict the search over \(\boldsymbol{\theta}\) by prespecifying a coarse candidate region.
This restriction’s purpose is to prevent the estimation routine from selecting parameter values that mechanically improve the objective function while implying economically implausible or numerically unstable equilibria.
Within this context, the parameter space serves as a preliminary admissible region that reflects both economic plausibility and computational feasibility.

Our estimation proceeds in two stages.
First, we perform a coarse grid search over a finite set of candidate values for each parameter and evaluate the objective function at all resulting combinations.
This step is used to identify promising regions of the parameter space and to screen out combinations that generate clearly implausible equilibrium outcomes, such as unreasonable vacancy rates, implausible qualifying score rankings, or unrealistic ordering between public and private colleges.\footnote{The gap between predicted and observed migrant flows, \(M_{ij}(\boldsymbol{\theta}) - M_{ij}^{\mathrm{obs}}\), is not employed as a direct criterion for determining the admissible coarse candidate region.}

Second, for each value of \(\lambda\), we select the best-performing coarse-grid point \(\check{\boldsymbol{\theta}}\) and run a local continuous optimization in the box
\(\check{\boldsymbol{\Theta}} = [0.90 \cdot \check{\boldsymbol{\theta}}, \, 1.10 \cdot \check{\boldsymbol{\theta}}]\).
In this step, \(\lambda\) is treated as a fixed grid parameter, and the continuous local optimization is applied only to \(\boldsymbol{\theta}\).
This choice is mainly computational: because \(\lambda\) changes the scale and curvature of the logit choice probabilities, joint quasi-Newton optimization over \(\lambda\) and \(\boldsymbol{\theta}\) can be numerically unstable in a nested problem where each objective-function evaluation requires solving an equilibrium.
Accordingly, the final estimate need not coincide with one of the coarse-grid values, and may lie outside the original coarse candidate set as long as it remains within the local optimization region.

\ref{ap:parameter_space} provides a more detailed explanation of how the coarse candidate region is chosen for each parameter and how economic plausibility and numerical admissibility are jointly used to discipline the search.

\begin{remark} \label{rm:scale}
In logit models, the scale of systematic utility relative to unobserved preferences requires a normalization (cf. Chapter 3 of \citealp{train2009}).
In this paper, we use a quasi-linear utility specification in which the coefficient on tuition $\tau_n$ is fixed at $-1$, providing a monetary scale normalization under which $\lambda$ can be estimated.
Under this normalization, $\lambda$ controls how strongly choices respond to systematic utility differences expressed in monetary units.
We therefore estimate $\lambda$ over the candidate grid described above.
\end{remark}

Standard errors are calculated following the tradition of the M estimator (cf. Chapter 22 of \citealp{hansen2022}).
The variance-covariance matrix $\widehat{\bm{V}}$ is calculated as follows:
\begin{align}
    \hat{\bm{V}} &=
    \hat{\bm{Q}}^{-1} \hat{\bm{\Omega}} \hat{\bm{Q}}^{-1} \label{eq:variance}, \\
    \hat{\bm{\Omega}} &=
    \frac{1}{I^2} \sum_{i=1}^I \sum_{j=1}^I \left(\frac{\partial}{\partial \bm\theta} \rho_{ij} (\hat{\bm{\theta}})\right) \left(\frac{\partial}{\partial \bm\theta} \rho_{ij} (\hat{\bm{\theta}})\right)^\top,  \notag \\
    \hat{\bm{Q}} &= \frac{1}{I^2} \sum_{i=1}^I \sum_{j=1}^I \frac{\partial^2}{\partial \bm\theta \partial \bm\theta^\top} \rho_{ij} (\hat{\bm{\theta}}). \notag
\end{align}
The score and Hessian in the preceding equations are evaluated numerically.

Table \ref{tab:estimate} reports the estimation results of the parameters.
Several features of the estimates are noteworthy.
First, the estimated moving-cost parameter, $\hat{\theta}_1=0.99$, lies inside the coarse candidate region rather than at its boundary.
This suggests that the model requires non-negligible inter-regional mobility frictions to match the observed enrollment flows, but does not rely on an extreme value of the moving cost parameter.
These mobility costs should be interpreted as generalized costs, including both pecuniary and nonpecuniary burdens of attending college away from home.

Second, the estimates of $\theta_2$ and $\theta_3$ should be interpreted jointly.
The relatively large value of $\hat{\theta}_2=451.93$ implies that higher qualifying scores generate a substantial effort-cost response, especially for lower-ability students.
Meanwhile, $\hat{\theta}_3=0.05$ implies a moderate peer-related gain from higher qualifying scores.
Thus, the estimated model places greater weight on the disciplining role of effort costs than on an excessively strong positive feedback from peer quality, which is also important for obtaining a numerically stable equilibrium.

Third, $\hat{\theta}_4=27.50$ indicates that differences in the education resource index play an important role in matching the observed spatial allocation of students.
As this estimate lies on the upper boundary of the local optimization region, it should not be interpreted as evidence that the coarse parameter range is a sharp theoretical bound.
Rather, it indicates that, within the local neighborhood searched in the outer loop optimization, the objective function still favors a relatively strong role for education resources.
For this reason, the economic meaning of $\theta_4$ is best understood through the equilibrium outcomes and counterfactual results, rather than as a stand-alone estimate of the monetary return to college resources.

Fourth, the estimated outside option is $\hat{u}_0=4.40$.
Because utility is measured in units of JPY 1 million under the quasi-linear specification, this value is comparable to observed household income levels in Japan.
For example, the 2023 Comprehensive Survey of Living Conditions reports that average household income was JPY 5.242 million and median household income was JPY 4.05 million.
Thus, the estimate of $u_0$ is economically plausible as the utility level associated with not enrolling in college.
The estimated scale parameter, $\hat{\lambda}=2.25$, suggests that systematic components of utility play a meaningful role in students' enrollment choices, while still allowing for idiosyncratic preference shocks.

Finally, the reported standard errors should be interpreted as local numerical curvature diagnostics rather than as fully robust measures of sampling uncertainty.
They are obtained by numerically evaluating the score and Hessian of the minimum-distance objective around the estimated parameter vector.
The very small values indicate that, within the local neighborhood used for the numerical differentiation, the objective function is highly sensitive to perturbations of the estimated parameters.
As each objective-function evaluation involves solving a numerical equilibrium and because the estimator is based on aggregate enrollment-flow moments, these standard errors should not be read as definitive confidence intervals.
Instead, they provide a local indication that the reported parameter vector is sharply determined within the maintained numerical procedure.

The calculation of numerical equilibrium is programmed by the statistical software {\ttfamily R}.
The optimization in \ref{step:opti:optimization} of the \ref{alg:outer} is carried out using the L-BFGS-B algorithm implemented through the {\ttfamily R} package \texttt{optimx}.
The parameters in Table \ref{tab:estimate} yield the minimum result of \eqref{eq:objective},
which is 642,239.
\begin{table}[t]
    \centering
    \caption{Estimates and Standard Errors of the Structural Parameters}
    \label{tab:estimate}
    \begin{tabular}{@{}lrr@{}}
        \toprule
        Parameter & Value & Standard error \\
        \midrule
        $\theta_1$ & $0.99$ & $1.14 \times 10^{-12}$ \\
        $\theta_2$ & $451.93$ & $1.99 \times 10^{-18}$ \\
        $\theta_3$ & $0.05$ & $8.27 \times 10^{-9}$ \\
        $\theta_4$ & $27.50$ & $5.06 \times 10^{-14}$ \\
        $u_0$ & $4.40$ & $2.06 \times 10^{-13}$ \\
        $\lambda$ & $2.25$ & - \\
        \bottomrule
    \end{tabular}
\end{table}

\subsection{Discussions about the Equilibrium} \label{subsec:discussion}
This subsection discusses the comparison between the equilibrium outcomes predicted by the model and the observed data. In this subsection and henceforth, we denote $M_{ij}$ as the abbreviation of $M_{ij}(\hat{\boldsymbol{\theta}})$.

\subsubsection{Model fit} \label{subsubsec:fitness}

\begin{figure}[tp]
    \begin{minipage}[b]{\textwidth}
        \centering
        \includegraphics[width=\linewidth]{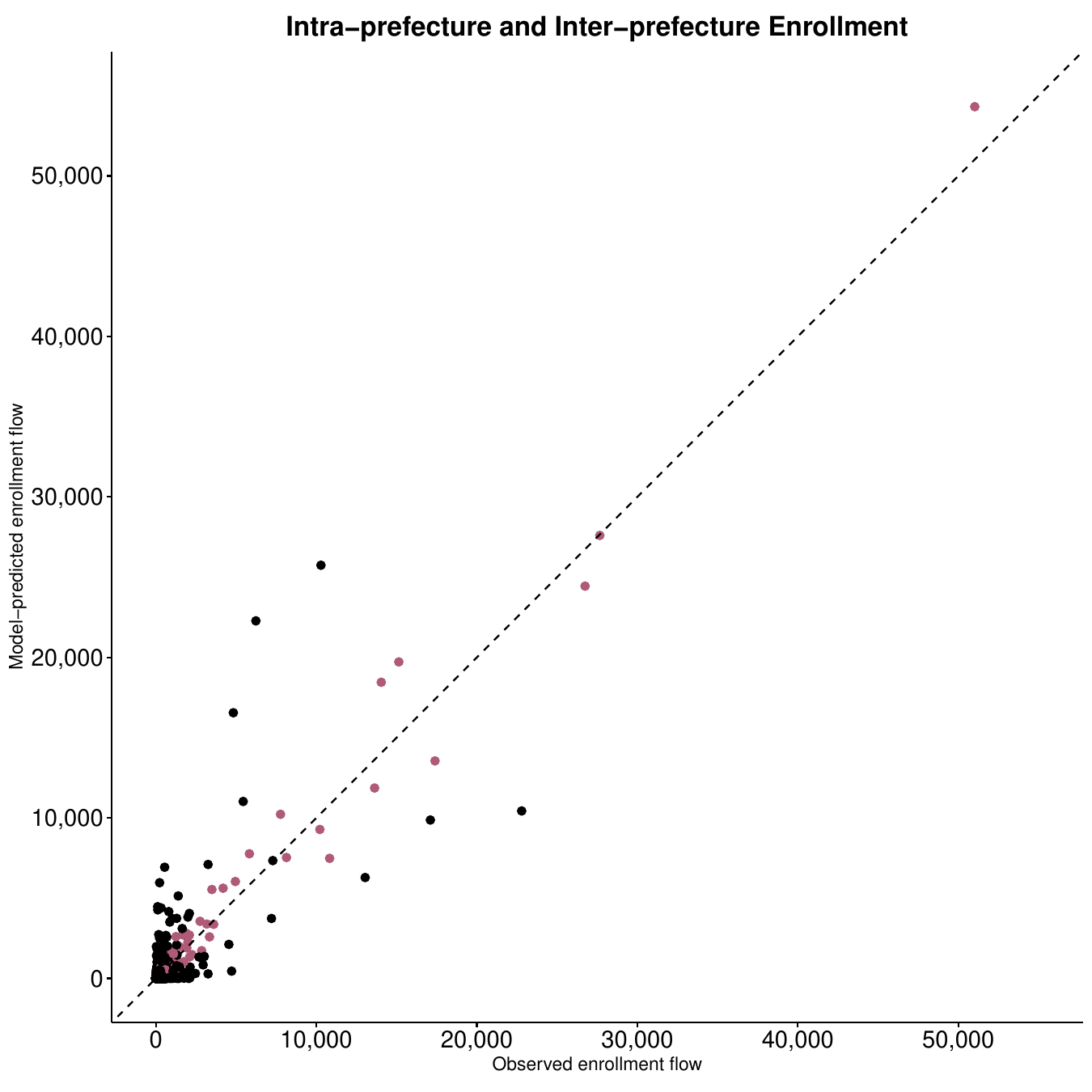}
        \captionsetup{width=.95\textwidth, font=footnotesize}
        \captionof{figure}{Comparison between observed enrollment flows $M^{\obs}_{ij}$ on the horizontal axis and model-predicted enrollment flows $M_{ij}$ on the vertical axis.
        Red points indicate intra-prefecture enrollment flows ($i=j$), while black points indicate inter-prefecture enrollment flows ($i \neq j$).
        The dashed line is the 45-degree line.
        The overall correlation coefficient is 0.908.
        The correlation coefficient is 0.987 for intra-prefecture enrollment flows and 0.668 for inter-prefecture enrollment flows.}
        \label{fig:ir_enrollment}
    \end{minipage}
\end{figure}

Figure \ref{fig:ir_enrollment} depicts the fitness by comparing the inter-regional enrollment between the observed data and model predictions.
In this figure, red points represent the intra-regional enrollment ($M_{ij}$ with $i = j$) and black points are inter-regional enrollment ($M_{ij}$ with $i \neq j$), respectively.

Notice that in Figure \ref{fig:ir_enrollment}, points close to the $45^\degree$ line are desirable. It appears that the fit for the intra-regional enrollments is better than the fit for the inter-regional enrollments.

\begin{figure}[tp]
    \begin{minipage}[b]{\textwidth}
        \centering
        \includegraphics[width=\linewidth]{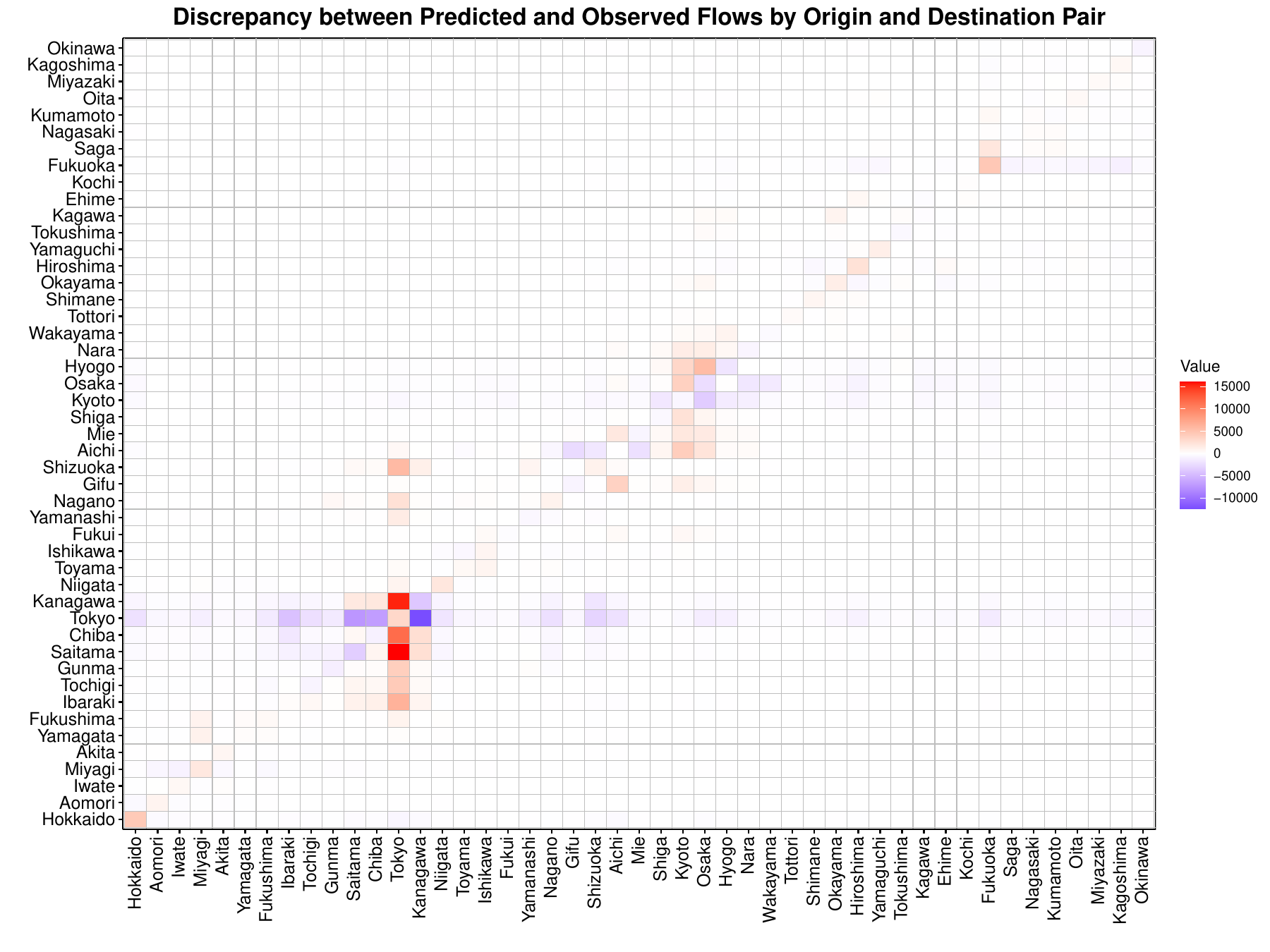}
        \captionsetup{width=.95\textwidth, font=footnotesize}
        \captionof{figure}{
        Difference of OD matrix $M_{ij} - M^{\obs}_{ij}$.
        Prefectures of the horizontal axis and vertical axis are the destination and origin, respectively.
        Blue grids indicate underestimation by the equilibrium and red grids signify overestimation.
        }
        \label{fig:od_diff}
    \end{minipage}
\end{figure}

To dig further, we present the origin-destination (OD) matrix of inter-regional enrollments differences ($M_{ij} - M^{\obs}_{ij}$) as depicted in Figure \ref{fig:od_diff}.
In this figure, prefectures on the horizontal axis and vertical axis are the destination and origin, respectively.
A blue grid indicates an underestimation by the model, while a red grid signifies an overestimation.

We highlight some facts indicated by Figure \ref{fig:od_diff}.
First, inflows to prefectures with large cities, such as Tokyo, Aichi, Kyoto, and Osaka, are overestimated, while outflows from these prefectures are underestimated.
This suggests that the model may not fully capture factors that discourage students from attending colleges in large cities, such as higher living costs.
Incorporating these factors into the model is left for future work.

Second, the bottom-left to the top-right diagonal of Figure \ref{fig:od_diff} represents the difference in intra-regional enrollment between model prediction and observed data.
Among these differences in intra-regional enrollment, Tokyo is overestimated, while Osaka and Hyogo are underestimated.
Aichi and Kyoto are a good fit.
Meanwhile, we observe insignificant overestimation of remote areas including Aomori, Iwate, Niigata, Yamaguchi, and Oita.

The OD pattern illustrated by Figure \ref{fig:od_diff} reveals a degree of spatial dependence in the estimation result of the residual of \eqref{eq:objective}.

\subsubsection{College enrollment rate and intra-regional enrollment rate} \label{subsubsec:college}

\begin{figure}[tp]
    \begin{minipage}[b]{\textwidth}
        \centering
        \includegraphics[width=\linewidth]{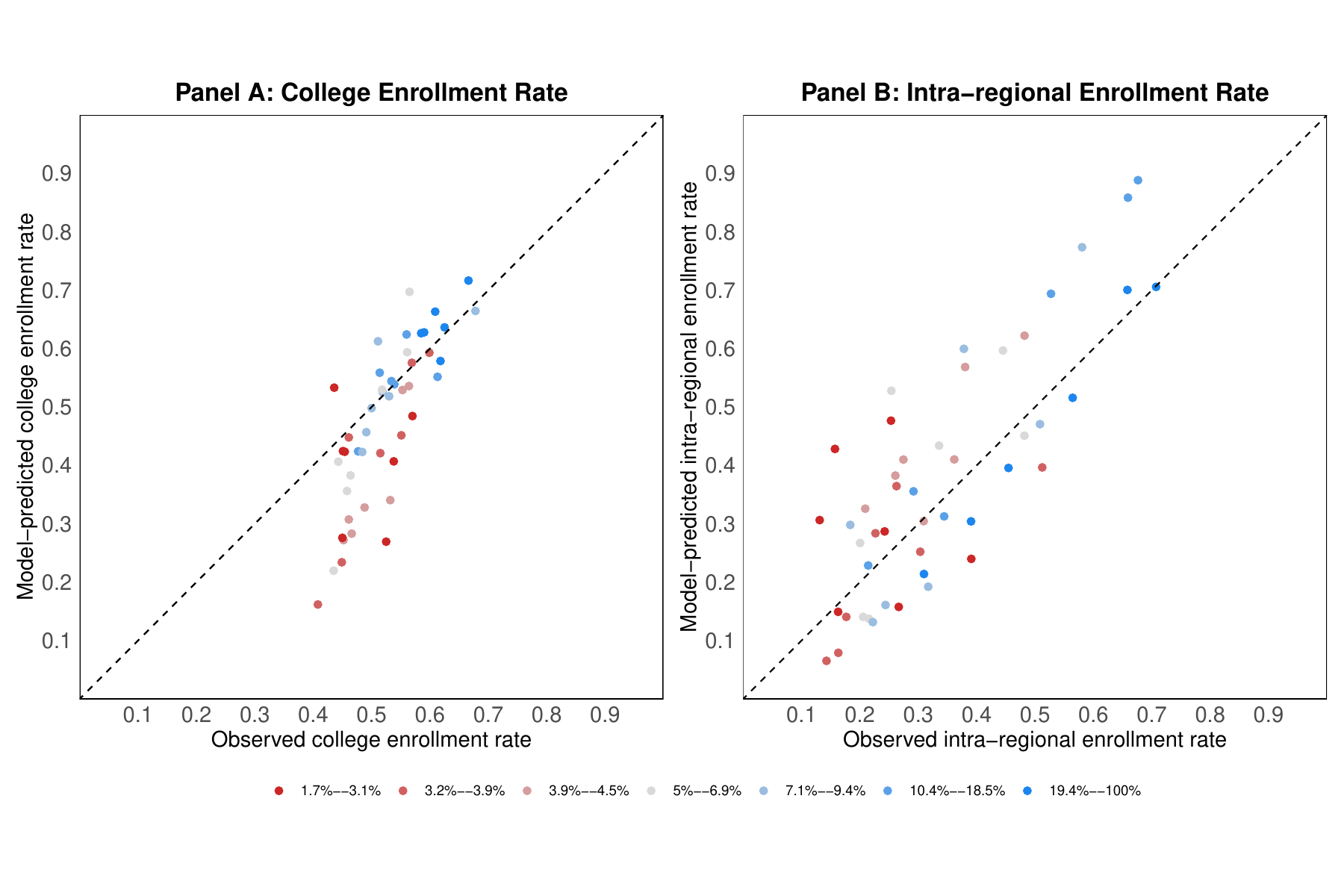}
        \captionsetup{width=.95\textwidth, font=footnotesize}
        \captionof{figure}{Panel A compares the college enrollment rate between the model prediction (vertical axis) and the observed data (horizontal axis).
        Panel B compares the intra-regional enrollment rate using the same axes.
        The dashed line is the $45^\degree$ line.
        Points below the dashed line represent underestimated prefectures, while points above the dashed line represent overestimated prefectures.
        Both axes use the same scale from 0 to 1.
        Points are colored by seven groups of GPP relative to the maximum GPP in the sample.
        From the lowest to the highest GPP group, the ranges are 1.7--3.1\%, 3.2--3.9\%, 3.9--4.5\%, 5.0--6.9\%, 7.1--9.4\%, 10.4--18.5\%, and 19.4--100.0\%, and the corresponding numbers of prefectures are 7, 7, 7, 7, 7, 6, and 6.
        The correlation coefficients are 0.792 in Panel A and 0.820 in Panel B.}
        \label{fig:equi_rate}
    \end{minipage}
\end{figure}

Panel A of Figure \ref{fig:equi_rate} compares the college enrollment rates across Japanese prefectures between the observed data on the horizontal axis and the model prediction on the vertical axis.
The dashed line represents the $45^\degree$ line.
Prefectures below the dashed line are underestimated by the model, while those above the dashed line are overestimated.
To avoid imposing arbitrary thresholds for urban and remote prefectures, the points are colored according to seven groups of prefecture-level Gross Prefecture Product\footnote{Data from the Japanese Cabinet Office, \url{https://www.esri.cao.go.jp/jp/sna/sonota/kenmin/kenmin_top.html}, in Japanese.} (GPP) relative to the maximum prefectural GPP in the sample.
The seven groups contain 7, 7, 7, 7, 7, 6, and 6 prefectures, respectively, from the lowest to the highest GPP group.

Panel A depicts that the model captures the broad cross-prefectural variation in college enrollment rates.
The points are concentrated around the $45^\degree$ line, and the correlation coefficient between the observed and predicted enrollment rates is 0.792.
As an additional diagnostic of model fit, the model overpredicts the outcome for 15 prefectures and underpredicts it for 32 prefectures.
The average prediction error is 0.047 among overpredicted prefectures and -0.094 among underpredicted prefectures.
These statistics suggest that the remaining discrepancies are somewhat asymmetric.
A closer inspection of the color pattern indicates that the model tends to underestimate college enrollment rates for several low-GPP prefectures, suggesting that the model does not fully capture the willingness to pursue higher education among students in economically smaller prefectures.
By contrast, high-GPP prefectures do not exhibit a clear tendency toward either underprediction or overprediction.
Therefore, the remaining prediction errors are not simply explained by a monotonic relationship between prefectural GPP and model fit.
Moreover, the color pattern indicates that prefectures with larger GPP tend to be located toward the upper-right part of the figure, while lower-GPP prefectures are more frequently located toward the lower-left part.
This suggests that the model preserves the broad ranking of prefectures by enrollment rates across economically different regions, although the figure should be interpreted as a descriptive diagnostic rather than as evidence of a causal relationship between GPP and enrollment outcomes.

Panel B of Figure \ref{fig:equi_rate} presents the intra-regional enrollment rate using the same visual structure.
The horizontal axis reports the observed intra-regional enrollment rate, while the vertical axis reports the model prediction.
The correlation coefficient is 0.820, indicating that the model also reproduces a substantial part of the cross-prefectural variation in students' tendency to enroll in colleges located within their home prefecture.
Compared with Panel A, the points in Panel B display larger dispersion around the $45^\degree$ line, implying that intra-regional enrollment patterns are harder to match exactly at the prefecture level.
Nevertheless, the figure shows that the model captures the overall positive association between observed and predicted intra-regional enrollment rates.
The color pattern does not indicate a clear systematic tendency for either low-GPP or high-GPP prefectures to be underpredicted or overpredicted.
Thus, while the model leaves non-negligible prefecture-level discrepancies in intra-regional enrollment rates, these discrepancies do not appear to be concentrated in a particular GPP group.

\subsubsection{Qualifying score} \label{subsubsec:qs}

\begin{figure}[tp]
    \begin{minipage}[b]{\textwidth}
        \centering
        \includegraphics[width=0.75\linewidth]{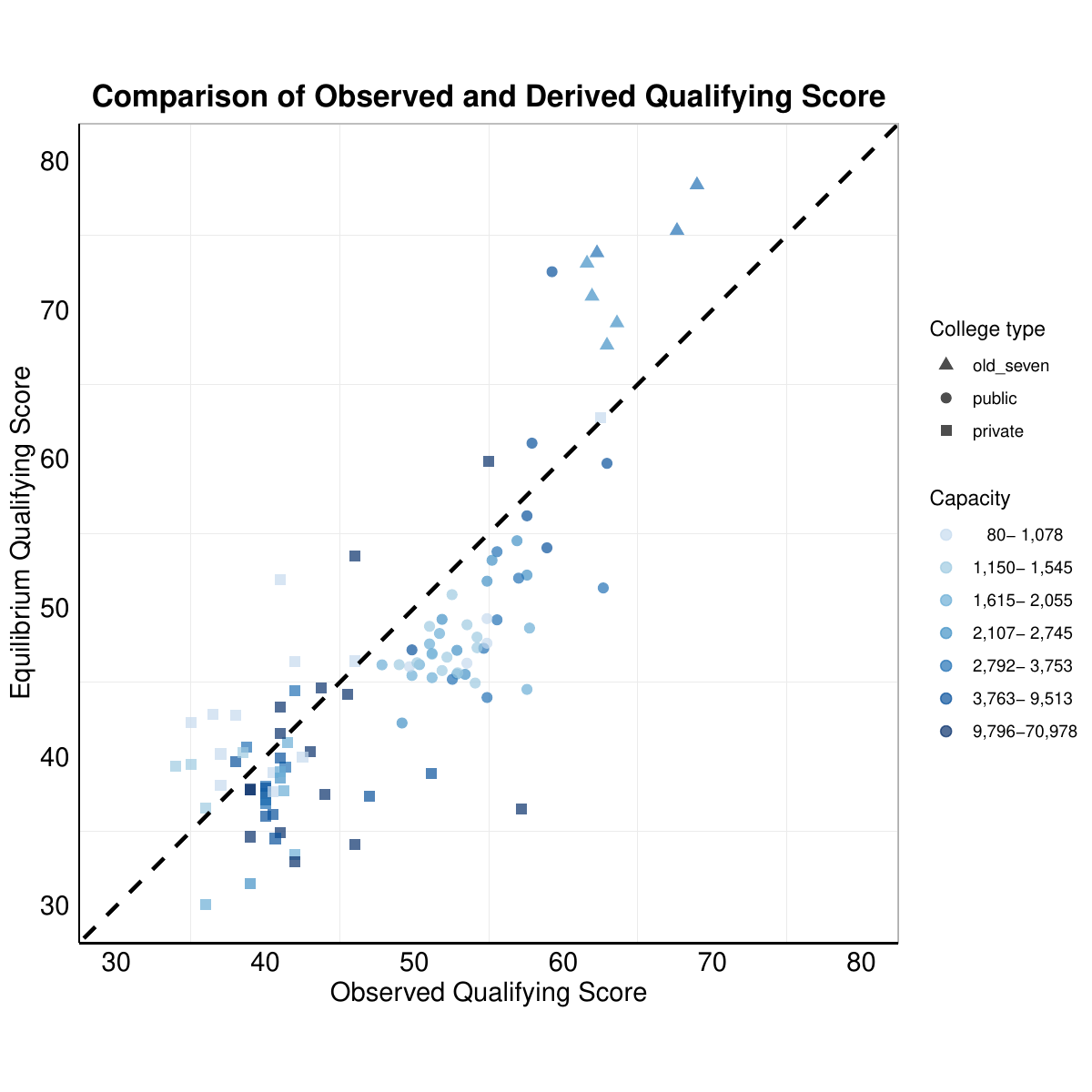}
        \captionsetup{width=.95\textwidth, font=footnotesize}
        \captionof{figure}{
        Comparison of the qualifying score between the model prediction (vertical axis) and the observed data (horizontal axis).
        The dashed line is the $45^\degree$ line.
        Points above the dashed line indicate that the model overpredicts the qualifying score, while points below the dashed line indicate underprediction.
        Marker shapes distinguish college types, and color intensity indicates enrollment capacity on a square-root scale.
        Correlation coefficient: 0.817 for all colleges, 0.853 for public colleges, and 0.526 for private colleges.
        }
        \label{fig:qs}
    \end{minipage}
\end{figure}

Figure \ref{fig:qs} depicts the comparison of the qualifying score between the observed data and model prediction.

The observed Hensachi and the equilibrium qualifying scores are positively correlated, with correlation coefficients of 0.817 for all colleges, 0.853 for public colleges, and 0.526 for private colleges, respectively.
These results support our model's validity in fitting the observed qualifying scores.

The model fits public colleges better than private colleges.
For public colleges, especially the high-ranking national universities, the points are concentrated along an upward-sloping pattern and remain relatively close to the $45^\degree$ line.
This result aligns with the model mechanism: Public colleges have more homogeneous tuition levels and their observed hierarchy is more closely aligned with education resources and capacity constraints, both of which are explicitly included in the model.
By contrast, private colleges display larger dispersion around the $45^\degree$ line.
The model tends to overpredict the qualifying scores of some private colleges with low observed Hensachi, including private college groups in prefectures such as Akita, Iwate, Aomori, Saga, Toyama, Nagano, and Yamaguchi.
These cases suggest that the model may attribute too much selectivity to colleges located in markets with limited nearby alternatives or relatively small capacity, while it does not fully capture private-college-specific factors such as reputation, program composition, local applicant pools, and within-group heterogeneity.
Conversely, the model underpredicts some large-capacity private college groups, such as those in Kyoto and Osaka.
For these colleges, large enrollment capacity mechanically lowers the market-clearing qualifying score in the model, whereas the observed Hensachi may remain high because of institutional reputation and selective programs within the aggregated private-college group.

Although the goodness of fit to observed qualifying scores is not included in the objective function \eqref{eq:objective}, the qualifying scores calculated by the model fit the data well. 
This result is consistent with the model's equilibrium mechanism, which links students' interregional mobility and college capacity to the determination of qualifying scores. 

\subsubsection{Regional disparities in accessibility to higher education} \label{subsubsec:education}

We can uncover the regional disparities in levels of higher education a little further by utilizing the equilibrium results.
This disparity stems from the lack of colleges suited to an individual's academic ability in her vicinity, particularly in remote regions.
Consequently, students in these regions face higher moving costs than students in regions with better access to higher education.
To measure this disparity by the equilibrium result, we define the average education quality $\mathcal{Q}$ and average qualifying score $\mathcal{E}$, respectively, as follows:
\begin{align*}
    \mathcal{Q}(a; \mathbf{e}^*, i) \coloneqq \sum_{n=0}^N P_{in}(a) \varphi(e^*_n, R_n) \quad \text{and} \quad \mathcal{E}(a; \mathbf{e}^*, i) \coloneqq \sum_{n=0}^N P_{in}(a) e^*_n,
\end{align*}
where we define $\varphi(e^*_0, R_0)=0$ and $e^*_0 = 0$.
Intuitively, for an individual who lives in prefecture $i$ with inherent ability $a$, $\mathcal{Q}$ and $\mathcal{E}$ measure the average quality of higher education and the corresponding average qualifying score, respectively.

\begin{figure}[tp]
    \begin{minipage}[b]{\textwidth}
        \centering
        \includegraphics[width=\linewidth]{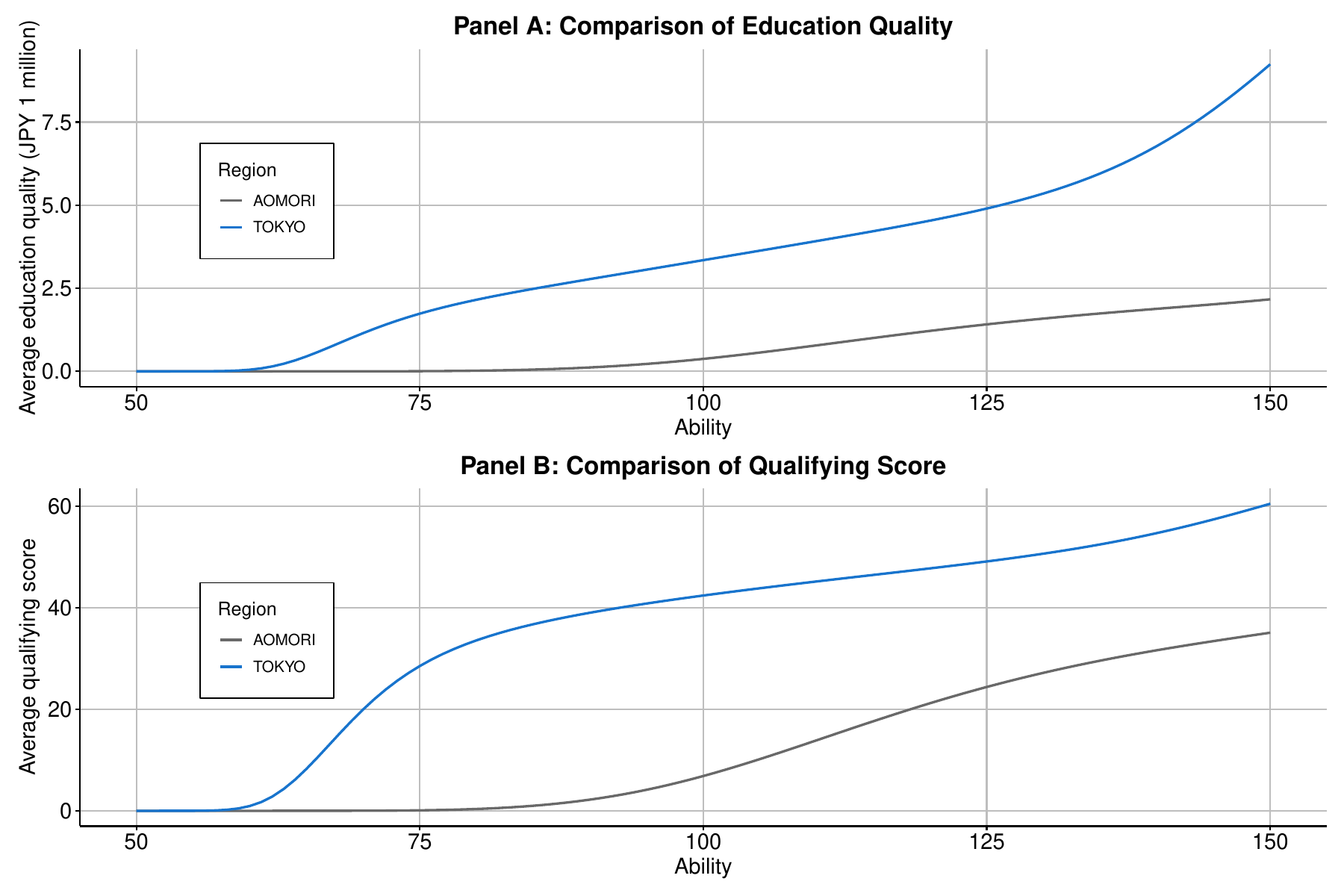}
        \captionsetup{width=.95\textwidth, font=footnotesize}
        \captionof{figure}{
        Examples of regional disparities in education quality (JPY 1 million).
        Blue solid line: Average education quality and qualifying score of Tokyo.
        Grey solid line: Average education quality and qualifying score of Aomori.
        }
        \label{fig:average_education}
    \end{minipage}
\end{figure}

Panel A of Figure \ref{fig:average_education} depicts the average education quality $\mathcal{Q}$ for individuals in an urbanized prefecture (represented by Tokyo) and a remote prefecture (represented by Aomori) over the support of ability $[\underline{a}, \overline{a}] = [50, 150]$.

We highlight some facts unraveled by Figure \ref{fig:average_education}.
The expectation of the average education quality $\mathcal{Q}$ for all high school graduates (integral of $\mathcal{Q}(a; \mathbf{e}^*, i)$ over the support of ability $[\underline{a}, \overline{a}]$ with respect to the density $g$) for Tokyo and Aomori are JPY 3.34 million and JPY 0.5 million, respectively.
This difference indicates a notable gap in education resources between the two regions.

First, Tokyo's $\mathcal{Q}$ exceeds that of Aomori over the range $[\underline{a}, \overline{a}]$, indicating that the hometown's location significantly impacts the quality of education that a student can receive.

Second, the minimum ability to attend college for Tokyo and Aomori are 58 and 78, respectively\footnote{In numerical calculation, we determine the minimum ability to attend college as the minimum ability to achieve $\mathcal{Q} \geq 0.01$. }.
This implies that lower-ability students in Tokyo have greater opportunities to attend higher education since there are many colleges within commuting distance.
That is, those from remote areas are required to have higher abilities than those in Tokyo in order to attend the same level of colleges.

Panel B of Figure \ref{fig:average_education} depicts the comparison of $\mathcal{E}$.
Resembling to that of $\mathcal{Q}$ in Panel A, we find out that the Tokyo's average qualifying score at each $a$ is higher than that of Aomori, indicating a notable gap in college access among urban and remote areas.

In sum, the hometown location significantly affects access to higher education.

\section{Welfare Costs of Disparity in Accessibility} \label{sec:counterfactual}

\begin{figure}[tp]
    \begin{minipage}[b]{\textwidth}
        \centering
        \includegraphics[width=\linewidth]{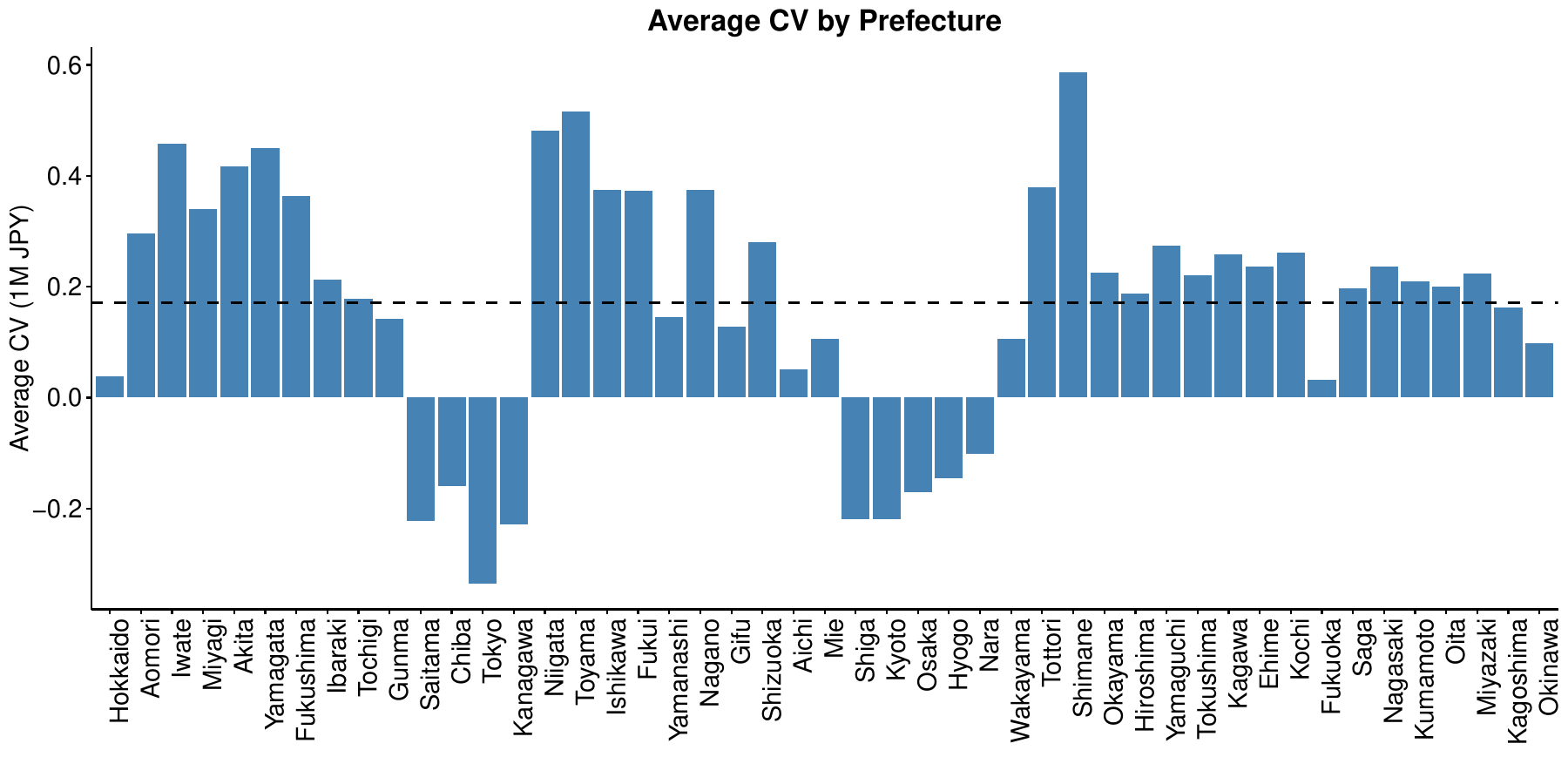}
        \captionsetup{width=.95\textwidth, font=footnotesize}
        \captionof{figure}{
        Average CV by prefecture in no-moving-cost counterfactual.
        Dashed black line: The mean value of average CV among all prefectures (JPY 1 million).
        Mean of average CV: 0.17.
        Standard deviation of average CV: 0.22. \\
        \textbf{Notice}: This figure reports region-specific average CVs, where each prefecture is given equal weight.
        Accordingly, the mean shown in the figure is an unweighted average across prefectures and should not be interpreted as the national per-capita welfare effect.
        }
        \label{fig:avg_cv}
    \end{minipage}
\end{figure}

This section quantitatively evaluates the welfare cost of disparity in accessibility to higher education.
The analysis is carried out by presuming a hypothetical case in which inter-regional moving costs are removed.

\subsection{Welfare Evaluation} \label{subsec:welfare}

We utilize compensating variation (CV) to measure the welfare change based on the discrete choice model
(cf. \citealp{small1981} or Chapter 2 of \citealp{small2007}).
Welfare of a student with ability $a$ in region $i$ is represented by the expected maximum utility, as follows:
\begin{equation}
E \left[ \max_{n \in \{0, \, 1, \, \dots, \, N\}} \left\{ V_{in}(a) + \varepsilon_{in} \right\} \right] = \frac{1}{\lambda} \log \left[ \sum_{n=0}^{N} \exp\left\{ \lambda V_{in}(a) \right\} \right].
\label{eq:cv_pre1}
\end{equation}
Suppose that the moving costs were eliminated to zero.
CV caused by this change for a student with ability $a$ in region $i$, denoted by $CV_i(a)$, satisfies the following equality:
\begin{equation}
    \log \left[ \sum_{n=0}^{N} \exp\left\{ \lambda V^0_{in}(a) \right\} \right] = \log \left[ \sum_{n=0}^{N} \exp\left\{ \lambda \left[ V^{\pc}_{in}(a)-CV_i(a) \right] \right\} \right]
    \label{eq:cv_pre2}
\end{equation}
where $V^0_{in}(a)$ and $V^{\pc}_{in}(a)$ are, respectively, the indirect utility levels in cases with and without moving costs, i.e.,
\begin{align*}
V^0_{in}(a) &= y_i - \tau_n-s_{in} - h(a,e^0_n) + \varphi(e^0_n,R_n), \\
V^{\pc}_{in}(a) &= y_i - \tau_n - h(a,e^{\pc}_n) + \varphi(e^{\pc}_n,R_n).
\end{align*}
$e^0_n$ and $e^{\pc}_n$ are the qualifying scores obtained by solving the equilibrium condition in \eqref{eq:market_clearing} for each of the two cases with and without moving costs, respectively.
From \eqref{eq:cv_pre2}, we have the formula to calculate compensating variations as follows:
\begin{equation}
CV_i(a) = \frac{1}{\lambda} \log \left[ \frac{\sum_{n=0}^{N} \exp\left\{ \lambda V^{\pc}_{in}(a) \right\}} {\sum_{n=0}^{N} \exp\left\{ \lambda V^0_{in}(a) \right\}} \right].
\label{eq:cv}
\end{equation}

\subsection{Welfare Costs of Disparity in Accessibility} \label{subsec:no_moving_cost}

We examine a scenario that spatial moving cost for any pair of regions is zero.
First, we assess the total CV of this policy counterfactual, which is calculated by
\begin{equation}
\sum_{i=1}^I {L}_i \int_{\underline{a}}^{\overline{a}} CV_i (a) {g}(a) \diff a = 21.25\ \text{(billion JPY)}.
\label{eq:cft:nsmc:total_cv}
\end{equation}
This implies that the welfare cost of disparity in accessibility to college education is JPY 21.25 billion.
To obtain quantitative insight, divide this number by the total prospective students, yields JPY 19.5 thousand per capita (in 2019).
This number represents the population-weighted average welfare effect per prospective student.
A policy instrument to realize this situation is providing a subsidy that is equal to $\theta_1 d_{ij}$ for each student.
The total amount of subsidy is calculated as follows
\begin{equation*}
    \sum_{i=1}^I \sum_{n=1}^N F^{\pc}_{in} \left(\sum_{j=1}^I \alpha_{nj} \theta_1 d_{ij}\right) = 2{,}423.57 \ \text{(billion JPY)}
\end{equation*}
where $F^{\pc}_{in}$ is the equilibrium number of students from region $i$ who attend college $n$ in the case without moving costs.
This amount could be interpreted as the cost of implementing the policy to completely remove the disparities in access to college education.
In this case,
the cost is too large to be practical.
Future research should deal with policy design that is effective but feasible.

Second, we evaluate the average CV of a specific prefecture $i$ through the expression:
\begin{equation*}
\int_{\underline{a}}^{\overline{a}} CV_i (a) {g}(a) \diff a.
\end{equation*}
Figure \ref{fig:avg_cv} visualizes the average CV across prefectures, with the dashed line representing the average CV for all prefectures.
Prefectures lacking access to higher education divisions benefit from significantly higher CV than the average, such as Aomori, Iwate, Akita, Tottori, and Shimane.
Among these prefectures, Shimane obtains the highest average CV, which is approximately JPY 0.59 million.

In contrast, prefectures in the Tokyo Metropolitan Area (Tokyo, Kanagawa, Chiba, and Saitama), the Kyoto-Osaka-Kobe Metropolitan Area (Kyoto, Osaka, and Hyogo), and the Nagoya Metropolitan Area (Aichi), experience a negative or near-zero CV gain.
Tokyo suffers the largest average CV loss, at approximately JPY 0.33 million.
This is because their privileged access to higher education divisions is taken away in the counterfactual scenario.
Neighbors of the aforementioned three metropolitan areas, such as Gunma, Mie, Shiga, Nara, and Wakayama, experience negative or below-average CV.

\begin{figure}[tp]
    \begin{minipage}[b]{\textwidth}
        \centering
        \includegraphics[width=\linewidth]{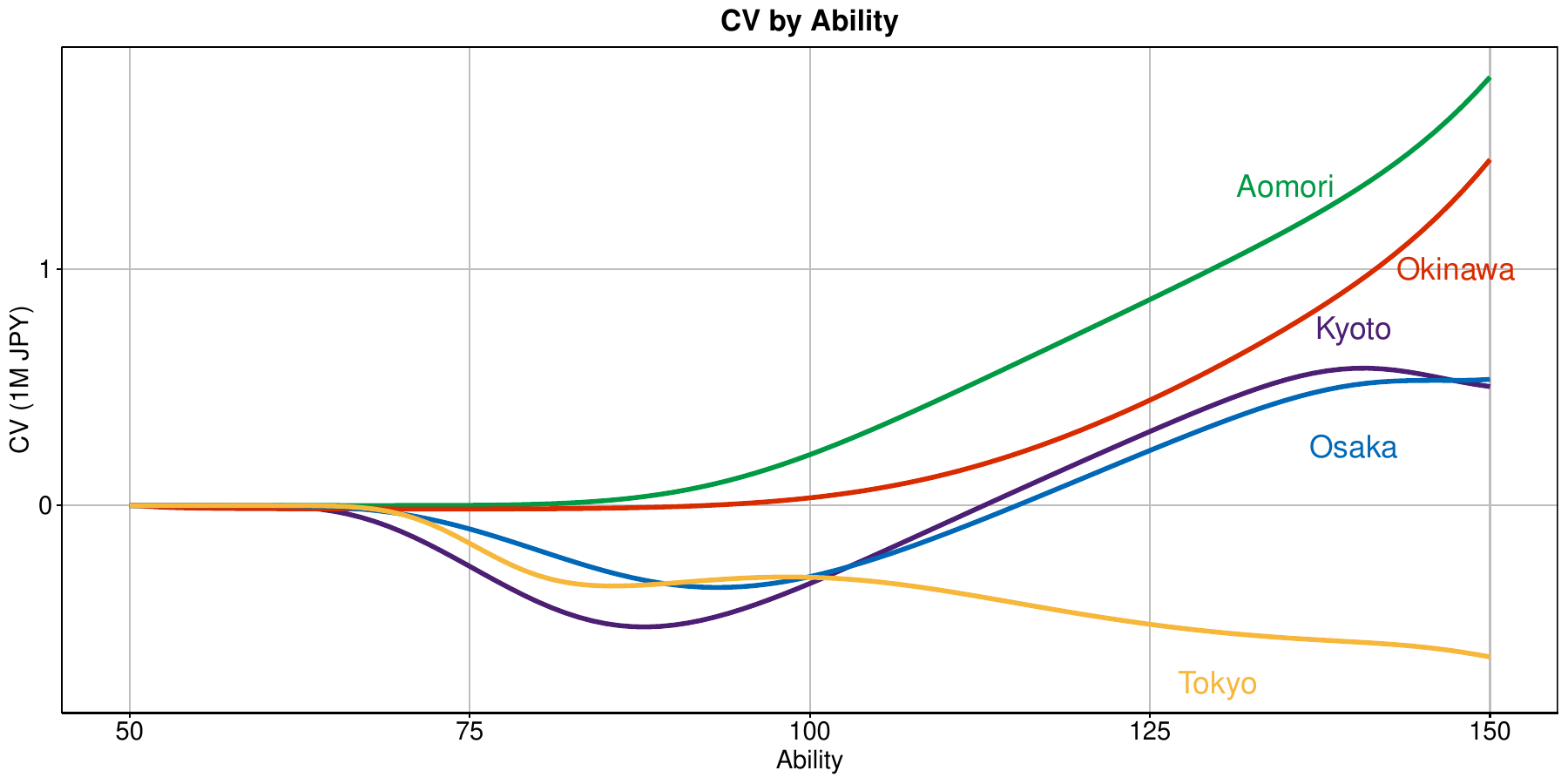}
        \captionsetup{width=.95\textwidth, font=footnotesize}
        \captionof{figure}{
        This figure plots $CV_i(a)$ for Tokyo, Kyoto, Osaka, Aomori, and Okinawa in the no-moving-cost policy counterfactual.
        }
        \label{fig:cv_segment}
    \end{minipage}
\end{figure}

Third, we assess the CV at each evaluated point of $a$ of five representative prefectures; Tokyo, Kyoto, Osaka, Aomori, and Okinawa.
Figure \ref{fig:cv_segment} illustrates the CV by ability for the selected prefectures.
Two remote prefectures, Aomori and Okinawa, yield non-negative CV across all evaluated ability levels.
In both prefectures, CV increases with ability, with Aomori showing a higher CV than Okinawa.

As urbanized prefectures, Kyoto and Osaka show similar CV patterns.
The CV of Kyoto (Osaka) starts to become negative at $a = 64$ ($a = 66$), and turn to positive at $a = 112$ ($a = 114$).

The CV of Tokyo exhibits an extreme case of losing advantage in college accessibility.
The trend of Tokyo's CV resembles those of Kyoto and Osaka but diverges at $a = 102$, dropping even lower thereafter.

\begin{figure}[tp]
    \begin{minipage}[b]{\textwidth}
        \centering
        \includegraphics[width=\linewidth]{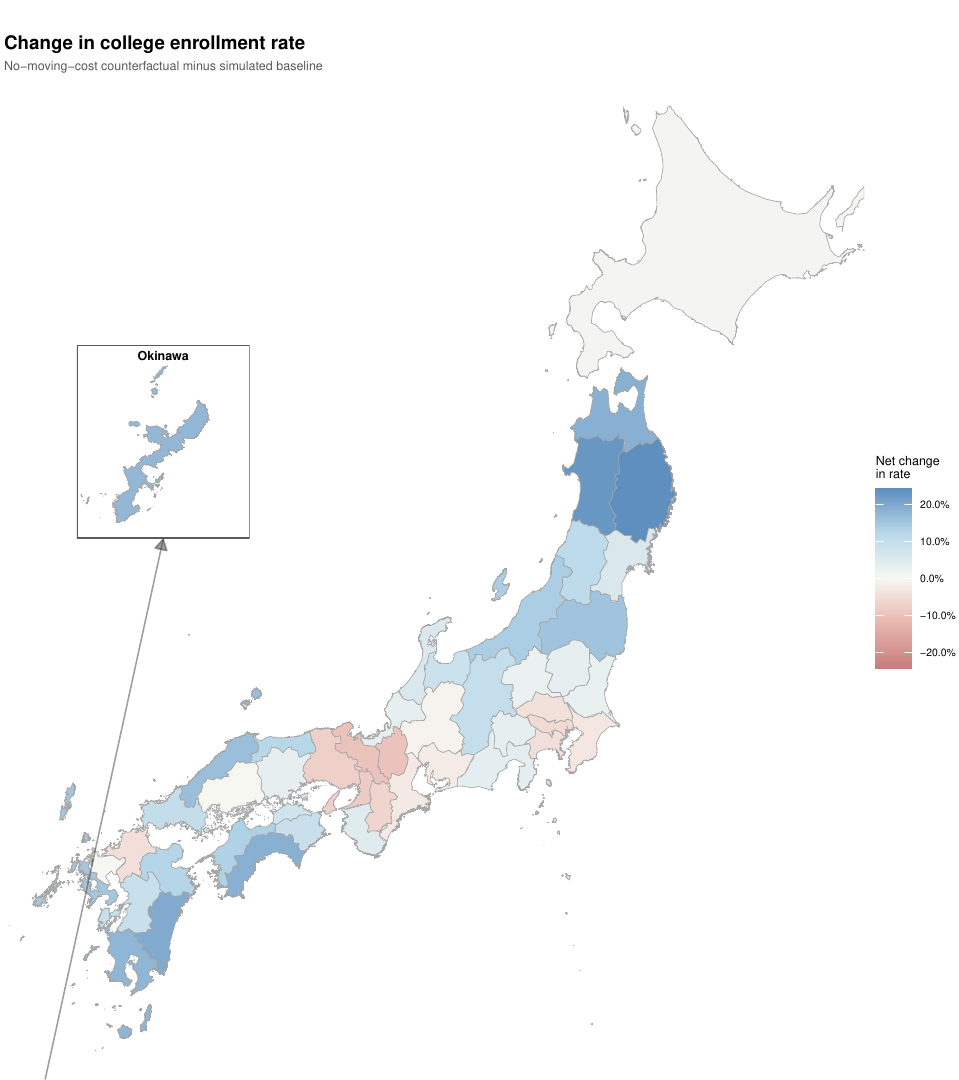}
        \captionsetup{width=.95\textwidth, font=footnotesize}
        \captionof{figure}{
        Change in prefecture-level college enrollment rates under the no-moving-cost counterfactual.
        The value for each prefecture is the counterfactual enrollment rate minus the simulated baseline enrollment rate.
        Blue indicates an increase and red indicates a decrease.
        The remaining variation reflects regional differences in utility bases that are not removed by eliminating moving costs.
        }
        \label{fig:pc_entrance}
    \end{minipage}
\end{figure}

Finally, we discuss the property in the college enrollment rates under this counterfactual scenario.
Figure \ref{fig:pc_entrance} depicts the net change in the college enrollment rate under the counterfactual.
Urban areas, where students already have relatively good access to colleges, experience declines in enrollment rates.
By contrast, remote prefectures with limited local access to colleges tend to experience increases in enrollment rates under the counterfactual.
When the inter-regional moving costs disappear, the disparity in access to higher education decreases drastically.
The range of the college enrollment rate is 16.2\%-71.7\% in the benchmark case and 33.5\%-66.3\% in the counterfactual case\footnote{Table \ref{tab:entrance_rate} in \ref{ap:subsec:entrance_rate} shows the college entrance rates of all prefectures in both the benchmark and counterfactual cases. }.

In sum, all prospective students in remote prefectures, regardless of their ability, benefit from the no-moving-cost counterfactual.
In urban prefectures, low-ability students experience welfare loss, while high-ability students gain welfare, though not as much as those in remote areas.
An exception among urban prefectures is Tokyo, where high-ability students lose more welfare than low-ability students.
Note that Tokyo has colleges with the highest quality.
Removing the access cost encourages students with higher ability from all regions to choose these colleges in Tokyo.
This increases competition for colleges in Tokyo, which leads to welfare loss for students from Tokyo owing to higher effort costs.
By contrast, high-ability students from Osaka and Kyoto benefit from the vanishing access cost to higher-quality colleges in Tokyo.

\section{Conclusion} \label{sec:conclusion}

This paper presents an equilibrium model of the higher education market, in which the cost of moving between regions significantly affects college attendance decisions of high school graduates.
The model is tailored to admission systems that rely on entrance exams, like those in Japan, South Korea, and China.
Students of different abilities make efforts to obtain higher scores on entrance exams.
Each college admits students with scores higher than the qualifying score up to its enrollment capacities.
Qualifying scores are endogenously determined as shadow prices of this capacity constraint.
We estimate the model's parameters using inter-regional student enrollment flows in Japan.
The model reproduces the spatial patterns of college attendance and qualifying scores.
It also quantitatively evaluates disparities in accessibility to higher education due to the uneven distribution of college locations.

Using the model, we conduct counterfactual simulations in which the moving costs are eliminated to zero.
The results demonstrate that eliminating moving costs significantly increases college attendance rates in remote regions.
Students from remote regions experience substantial welfare gains, whereas students from urban regions may experience losses.
Due to the presence of negative gains like this, the aggregate welfare gain for the whole country is moderate.

Eliminating regional disparities is not easy since moving costs include non-monetary elements.
Even if it were feasible through government policies, our simulations show that the cost of these policies would far exceed the benefits.
Future studies should consider feasible and effective policies, such as providing a subsidy for a portion of the moving costs.
Our model would be useful for examining various alternative subsidy designs, such as the optimal subsidy rate, and differentiated subsidy rates for different regions.

Future studies should also improve the model formulation by relaxing some of the simplifying assumptions.
Among them, we assume that students pass the exam by choosing the level of effort that will result in scores that match the qualifying scores exactly.
It would be more realistic to introduce uncertainty as to whether students will pass the exam, and describe more concrete steps in the application and entrance process.\footnote{In this direction, \cite*{qin2025admission} develop a Bayesian equilibrium model of China's higher education market that incorporates admission uncertainty. In their framework, students make application and enrollment-related decisions without fully observing the actual qualifying scores of universities.}

Another important direction concerns the empirical and econometric methodology.
The present paper estimates the structural parameters through a nested equilibrium-and-matching procedure, in which the equilibrium qualifying scores are solved numerically for each candidate parameter vector and the resulting inter-regional enrollment flows are matched to the observed flows.
This approach is suitable for the aggregate data currently available, but several methodological issues remain.
Future studies could develop more efficient minimum distance (MD) or simulated method of moments (SMM) estimators with explicit weighting matrices, examine the sensitivity of the estimates to the assumed ability distribution and numerical equilibrium selection, and construct confidence intervals that account for both sampling variation in the observed enrollment flows and numerical uncertainty from the equilibrium computation.
Richer data on individual applications, admissions, and enrollment decisions would also make it possible to identify more detailed heterogeneity in moving costs, college preferences, and application behavior.

\section*{Acknowledgements}
Earlier versions of this paper were presented at the Applied Regional Science Conference in Osaka, Japan, and the seminar at Kyoto University. 
We thank the conference participants and Yuta Kuroda, Yasusada Murata, Yoshihiko Nishiyama, Takatoshi Tabuchi, Keisuke Takano for their helpful comments and suggestions. 
This research was partially supported by JSPS Grants-in-Aid for Scientific Research (Number 20H01494).

\section*{Declaration of generative AI and AI-assisted technologies}

During the preparation and final review of this manuscript, the authors used OpenAI’s ChatGPT and Codex for three supporting tasks. 
First, these tools assisted with proofreading and language review, including the identification of typographical errors and improvements in fluency, clarity, and consistency of academic style. 
Second, they assisted in reviewing code previously developed by the authors and checking its consistency with the methods described in the manuscript, including possible implementation errors, conceptual discrepancies, and methodological steps not adequately reflected in the manuscript. 
Third, they assisted with adjustments to the color schemes, visual style, and layout of figures originally developed by the authors. 
These presentation adjustments did not change the underlying data, calculations, or reported results. 
The authors reviewed and verified the AI-assisted suggestions and edits as needed and take full responsibility for the content of the manuscript. 

\clearpage

\bibliographystyle{tandfx}
\bibliography{references}

@book{Obunsha,
  editor = {Obunsha},
  title = {Keisetsu Jidai, 8-gatsu Rinji Zokan, Zenkoku Daigaku Naiyo Annai-go (Keisetsu Jidai, August, in Japanese)},
  year = {2020},
  publisher = {Obunsha}
}

@article{kuroha1995, 
  author = {Kuroha, Ryoichi}, 
  title = {Nihon ni okeru 1990-nendai no daigaku kaikaku (Japanese Higher Education Reform in 1990's, in Japanese)}, 
  journal = {Research in Academic Degrees}, 
  number = {3},
  pages = {1-41},
  year = {1995}, 
  month = {06}, 
  issn = {0919-6099}, 
  resource_type = {departmental bulletin paper}, 
  note = {in Japanese} 
}

@article{Kobayashi2006,
  author = {Kobayashi, Masayuki},
  title = {Kotokyoiku no Chiho Bunsan-ka Seisaku no Kensho (An Analysis of Geographical Dispersion Policy of Higher Education, in Japanese)},
  journal = {Kotokyoiku Kenkyu (Research on Higher Education, in Japanese)},
  volume = {9},
  pages = {101-120},
  year = {2006}
}

@book{Hasegawa2016,
  author = {Hasegawa, Makoto},
  title = {Daigaku Zennyu Jidai ni Okeru Shinro Ishiki to Shinro Keisei -- Naze Yonen-sei Daigaku ni Shingaku Shinainoka (Why Do They Go to Four-Year Universities?, in Japanese)},
  publisher = {Minerva Press},
  year = {2016}
}

@book{Yano2015,
  author = {Yano, Masakazu},
  title = {Daigaku no Joken -- Taishu-ka to Shijo-ka no Keizai Bunseki (Conditions of Universities, Economic Analysis on Popularization and Marketization, in Japanese)},
  publisher = {Tokyo University Press},
  year = {2015}
}

@article{Yamamoto1979,
  author = {Yamamoto, Shinichi},
  title = {Daigaku Shingaku Kibo-ritsu Kitei Yoin no Bunseki (An Analysis on Factors Determining the Rate of University-Bounded Students, in Japanese)},
  journal = {Kyoiku-Shakaigaku Kenkyu (Research in Educational Sociology, in Japanese)},
  volume = {34},
  pages = {93-103},
  year = {1979}
}

@article{epple2006,
  author = {Epple, Dennis and Romano, Richard and Sieg, Holger},
  title = {Admission, Tuition, and Financial Aid Policies in the Market for Higher Education},
  journal = {Econometrica},
  volume = {74},
  number = {4},
  pages = {885-928},
  year = {2006}
}

@incollection{sacerdote2011,
  author = {Sacerdote, Bruce},
  title = {Peer Effects in Education: How Might They Work, How Big Are They and How Much Do We Know Thus Far?},
  booktitle = {Handbook of the Economics of Education 3},
  editors = {Eric A. Hanushek, Stephen Machin, Ludger Woessmann},
  publisher = {Elsevier},
  pages = {249-277},
  year = {2011}
}

@article{small1981,
  author = {Small, Kenneth A. and Rosen, Harvey S.},
  title = {Applied Welfare Economics with Discrete Choice Models},
  journal = {Econometrica},
  volume = {49},
  number = {1},
  pages = {105-130},
  year = {1981}
}

@article{Muta1986,
  author = {Muta, Hiromitsu},
  title = {Daigaku \& Tandai Shingaku ni Tomonau Chiiki-kan Ido no Jikeiretsu Bunseki (Time-Series Analysis of Interregional Movements Related to Entrance into Universities and Colleges, in Japanese)},
  journal = {Daigaku Ronshu},
  volume = {16},
  pages = {181-197},
  year = {1986}
}

@article{Asia,
  author = {{Asian Growth Research Institute}},
  title = {Daigaku Shingaku ni Tomonau Todofukenkan Jinko Ido (Movements among Prefectures Related to Entrance into Universities, in Japanese)},
  year = {2017},
  note = {16-08}
}

@article{Fillmore2023,
    author = {Fillmore, Ian},
    title = "{Price Discrimination and Public Policy in the US College Market}",
    journal = {The Review of Economic Studies},
    volume = {90},
    number = {3},
    pages = {1228-1264},
    year = {2022},
    month = {07},
    issn = {0034-6527},
    doi = {10.1093/restud/rdac051}
}

@article{FU2014,
 ISSN = {00223808, 1537534X},
 URL = {http://www.jstor.org/stable/10.1086/675503},
 author = {Chao Fu},
 journal = {Journal of Political Economy},
 number = {2},
 pages = {225--281},
 publisher = {The University of Chicago Press},
 title = {Equilibrium Tuition, Applications, Admissions, and Enrollment in the College Market},
 urldate = {2023-11-06},
 volume = {122},
 year = {2014}
}

@article{EPPLE2017,
title = {A general equilibrium analysis of state and private colleges and access to higher education in the U.S.},
journal = {Journal of Public Economics},
volume = {155},
pages = {164-178},
year = {2017},
issn = {0047-2727},
doi = {https://doi.org/10.1016/j.jpubeco.2016.09.003},
url = {https://www.sciencedirect.com/science/article/pii/S004727271630130X},
author = {Dennis Epple and Richard Romano and Sinan Sarpça and Holger Sieg}
}

@article{FU2022,
title = {Students’ heterogeneous preferences and the uneven spatial distribution of colleges},
journal = {Journal of Monetary Economics},
volume = {129},
pages = {49-64},
year = {2022},
issn = {0304-3932}, 
author = {Chao Fu and Junjie Guo and Adam J. Smith and Alan Sorensen}
}

@book{hansen2022,
  title={Econometrics},
  author={Hansen, Bruce},
  isbn={9780691235899},
  lccn={2021049283},
  url={https://books.google.co.jp/books?id=VaRcEAAAQBAJ},
  year={2022},
  publisher={Princeton University Press}
}

@book{small2007,
  author    = {Kenneth A. Small and Erik T. Verhoef},
  title     = {The Economics of Urban Transportation},
  publisher = {Routledge},
  edition   = {2nd},
  year      = {2007},
  month     = {October 4},
  day       = {4},
  pages     = {296},
  isbn-10   = {0415285151},
  isbn-13   = {978-0415285155},
}

@misc{cook2021competing,
  title={Competing Campuses: Equilibrium Prices, Admissions, and Undergraduate Programs in US Higher Education},
  author={Cook, Emily},
  year={2021},
  month={November 23},
  note={Available at SSRN: \url{https://ssrn.com/abstract=3970159}}
}

@article{Yonezawa2023,
author = {Yonezawa, Akiyoshi},
title = {Japan's Higher Education Policies under Global Challenges},
journal = {Asian Economic Policy Review},
volume = {18},
number = {2},
pages = {220-237},
doi = {https://doi.org/10.1111/aepr.12421},
url = {https://onlinelibrary.wiley.com/doi/abs/10.1111/aepr.12421},
eprint = {https://onlinelibrary.wiley.com/doi/pdf/10.1111/aepr.12421},
year = {2023}
}

@article{Ishimaru2025,
author = {Ishimaru, Shoya},
title = {Geographic Mobility of Youth and Spatial Gaps in Local College                         and Labor Market Opportunities},
journal = {Journal of Labor Economics},
volume = {43},
number = {4},
pages = {1251-1294},
year = {2025},
doi = {10.1086/730542},

URL = { 
    
        https://doi.org/10.1086/730542
    
    

},
eprint = { 
    
        https://doi.org/10.1086/730542
    
    

}
}

@book{train2009, 
place={Cambridge}, 
edition={2}, 
title={Discrete Choice Methods with Simulation}, 
publisher={Cambridge University Press}, 
author={Train, Kenneth E.}, year={2009}}

@incollection{cai2017,
  author    = {Cai, Yuzhuo and Yan, Feng},
  title     = {Higher Education and University},
  booktitle = {Handbook of Chinese Education},
  editor    = {Morgan, W. J. and Gu, Qiang and Li, Fang},
  publisher = {Edward Elgar},
  year      = {2017},
  pages     = {169-193},
}

@article{Hafalir2018,
title = {College admissions with entrance exams: Centralized versus decentralized},
journal = {Journal of Economic Theory},
volume = {176},
pages = {886-934},
year = {2018},
issn = {0022-0531},
doi = {https://doi.org/10.1016/j.jet.2018.05.009},
url = {https://www.sciencedirect.com/science/article/pii/S0022053118301856},
author = {Isa E. Hafalir and Rustamdjan Hakimov and Dorothea Kübler and Morimitsu Kurino}
}

@unpublished{Fabre2023,
  author       = {Fabre, Ana{\"i}s},
  title        = {The Geography of Higher Education and Spatial Inequalities},
  year         = {2023},
  note         = {Job Market Paper / Working Paper},
}

@article{Gibbons2012,
  author       = {Gibbons, Stephen and Vignoles, Anna},
  title        = {Geography, Choice and Participation in Higher Education in England},
  journal      = {Regional Science and Urban Economics},
  year         = {2012},
  volume       = {42},
  number       = {1-2},
  pages        = {98--113},
}

@article{Newbold2015,
  author       = {Newbold, K. Bruce and Brown, W. Mark},
  title        = {The Urban--Rural Gap in University Attendance: Determinants of University Participation Among Canadian Youth},
  journal      = {Journal of Regional Science},
  year         = {2015},
}

@article{Redding2017,
  author  = {Redding, Stephen J. and Rossi-Hansberg, Esteban},
  title   = {Quantitative Spatial Economics},
  journal = {Annual Review of Economics},
  year    = {2017},
  volume  = {9},
  number  = {1},
  pages   = {21--58},
  doi     = {10.1146/annurev-economics-063016-103713}
}

@article{ahlfeldt2015,
    author = {Ahlfeldt, Gabriel M. and Redding, Stephen J. and Sturm, Daniel M. and Wolf, Nikolaus},
    title = {The Economics of Density: Evidence From the Berlin Wall},
    journal = {Econometrica},
    volume = {83},
    number = {6},
    pages = {2127-2189},
    doi = {https://doi.org/10.3982/ECTA10876},
    url = {https://onlinelibrary.wiley.com/doi/abs/10.3982/ECTA10876},
    year = {2015}
}

@misc{qin2025admission,

  author       = {Qin, Lei and Tang, Xiwen},

  title        = {An Equilibrium Analysis on China's Higher Education Market under Admission Uncertainty},

  year         = {2025},

  howpublished = {Presented at the 39th Annual Meeting of the Applied Regional Science Conference}

}

\clearpage

\appendix

\refstepcounter{section}
\setcounter{subsection}{0}
\section*{Appendix \thesection\quad Higher Education Market in Japan} \label{ap:japan}

This section provides an overview of the higher education market and admission system in Japan and compares them with those in other countries.

\subsection{The Japanese Colleges} \label{ap:subsec:college}

\textbf{The public and private colleges.}
Different from the circumstances of the United States higher education market, Japan's top-tier higher education institutions are predominantly public colleges.
These public colleges are often comprehensive and receive substantial research funding from the Japanese government compared to their private counterparts.
In general, public colleges tend to be more selective on average and often have stronger research-oriented institutional resources than private colleges.

Among the public colleges, the \textit{former imperial universities} have prestigious status and fame in Japan.
This terminology refers to the seven Imperial Universities established in Japan before World War \uppercase\expandafter{\romannumeral 2}, including University of Tokyo (in Tokyo), Kyoto University (in Kyoto), Tohoku University (in Miyagi), Kyushu University (in Fukuoka), Hokkaido University (in Hokkaido), Osaka University (in Osaka), and Nagoya University (in Aichi).

Private colleges also play an important role in Japan's higher education market.
As private-sector suppliers of higher education, private colleges offered 79.4\% of Japan's higher education seats in 2019.
This figure highlights their substantial contribution to Japan's human capital formation.
The private sector is highly heterogeneous.
Prestigious private colleges, such as Keio University and Waseda University, attract top students from Japan and overseas and provide high-quality education. 

\textbf{The uneven distribution of colleges in Japan.}
Right after World War \uppercase\expandafter{\romannumeral 2}, there were only 11,978 students and 12 colleges in Japan in 1948.
We witness rapid growth in student body after this period.
In 1953, the number of students increased to about 450,000, and this figure continued to reach 2,000,000 and 2,920,000 in 2000 and 2020, respectively.
Correspondingly, the number of colleges increased simultaneously as well.
There were about 400 colleges in 1973, and this number increased to 600 in 1988.

However, the colleges are not evenly distributed across prefectures.
Specifically, there are more higher education divisions in well-developed prefectures than those in underdeveloped prefectures.
Panels C and D of Figure \ref{fig:prefecture_comparison} depict the Gross Prefecture Product (GPP) and the number of colleges by prefectures in 2019, respectively.
By comparing these figures, we conclude that well-developed prefectures (in terms of GPP) possess a larger body of higher education divisions, especially for the Tokyo Metropolitan Area, Kyoto-Osaka-Kobe Metropolitan Area, Nagoya Metropolitan Area, and some local economic centers like Hiroshima, Fukuoka, Miyagi, and Hokkaido.

This uneven distribution leads to not only the disparity in chances of attending colleges across prefectures but also the movements of students from remote prefectures to urban ones.
To mitigate this phenomenon, the Japanese government took measures after the 1970s, trying to promote the geographical dispersion of colleges' locations.
This government-led dispersion policy continued for 20 years until the 1990s and was officially terminated at the 5th National Plan on Higher Education in Japan in 1997.
Researchers have examined whether this policy was effective, but they have not arrived at a consensus \citep{kobayashi2006}.
The details of this policy in the 1990s are summarized in \cite{kuroha1995}.

\subsection{The Japanese College Entrance Exam} \label{ap:subsec:entrance}

About 50\% of high school graduates choose to attend college every year in Japan. Students pursuing higher education look for a college suitable for their needs and abilities and, in general, undergo a paper-based entrance exam organized by the college they choose. Colleges only recruit students whose exam scores are higher than a threshold as enrollees.

This threshold is phrased as the minimum qualifying score in this paper. The most widely used proxy for Japanese colleges' threshold is the \textit{`Hensachi'}. Hensachi is a z-score that evaluates the gap from the average score:
\begin{align*}
    \text{Hensachi} \coloneqq 10 \cdot \frac{\text{individual score} - \text{average score}}{\text{standard deviation}} + 50.
\end{align*}
Although Hensachi is not theoretically bounded, published values for Japanese colleges are typically concentrated within a limited range.
In practice, the upper end rarely exceeds 80, whereas the lower end is often bottom-coded (e.g., reported as 35.0 for values below 37.5) or labeled BF (“border-free”).
For computational purposes, we therefore restrict Hensachi to the interval $(30,80)$ in our analysis.

The Hensachi plays an important role in determining the minimum score requirement for enrollment and representing the ranking of colleges.
For example, as one of the highest requirements of Hensachi, the Faculty of Medicine, Kyoto University asks for a Hensachi of 72.5 (about top 1.22\%) for enrollment.
As top-tier colleges among public and private colleges, both Faculty of Economics, Hitotsubashi University and Faculty of Economics, Keio University require a Hensachi of 65 (about top 6.68\%).

Earning higher scores to enroll the top-tier college takes great effort.
A common choice for Japanese high school students is to attend preparatory school to raise their scores.
Preparatory schools in Japan are specifically designed to prepare students for university entrance examinations.
These institutions play a crucial role in the Japanese educational system, especially for students aiming to gain admission to competitive and prestigious colleges.

The primary goal of preparatory schools is to enhance students' scores on college entrance exams. They offer intensive courses on a wide range of subjects relevant to these exams. Besides subject-specific knowledge, preparatory schools also teach test-taking strategies, time management skills, and efficient study techniques.

The attendance fee of preparatory schools in Japan varied widely depending on the type of school, its location, the comprehensiveness of the programs offered, and whether the student attended full-time or part-time. For part-time programs or specific subject courses, students might pay between JPY 100,000 and JPY 300,000 per year. Full-time programs could cost significantly more, with fees ranging from approximately JPY 800,000 to JPY 1,500,000 per year.

The studying cost $h(a, e_n)$ we introduce in \eqref{eq:budget} stems from this attendance fee for preparatory schools. We believe this studying cost function mirrors the competitive nature of the college entrance exam phenomenon in Japan.

\subsection{Applicability of the Model to the China Higher Education Market} \label{ap:subsec:comparison}

In this subsection, we discuss the applicability of the model in this paper to the case of the higher education market in China.
In China, the government, instead of colleges themselves, is responsible for conducting a countrywide paper-based college entrance exam (named \textit{`Gaokao'} in Chinese) every year.
Almost all provinces of China use the same exam made and marked by the Ministry of Education of the People's Republic of China, except a few provinces or municipalities are responsible for the exam of their own.
The outline of the Gaokao system is summarized by \cite{cai2017}.

The enrollment of colleges that provide undergraduate-level higher education in China is almost exclusively determined by the Gaokao score.
Similar to the case of Japan, attending preparatory schools is also a popular choice in China to raise the Gaokao score.
The attendance fee for preparatory schools in China depends on the quality, subjects, class location, and class size.
For example, an onsite class with 10-20 students may cost 100-200 CNY per student per hour, and the attendance fee for 1-1 online classes can be 300-600 CNY per hour.
Thus, the specification of studying cost $h(a, e_n)$ in our model also suits the case of China.

Since the Gaokao score is a uniform and authorized measure, it serves as a better proxy for colleges' minimum qualifying score than the Hensachi of Japan's case. However, colleges in China have enrollment capacities for each province which are determined by the government, while colleges in Japan do not impose restrictions based on the students' prefectures of birth. This results in differences of minimum qualifying score across provinces: The qualifying score for science students of Peking University in 2019 for Hebei Province and Shanxi Province are 696 and 665, respectively\footnote{In these two provinces, the total score of Gaokao is 750. }, although these two provinces used the same test papers.

In summary, the model proposed in this paper is also suitable for the context of China with some specific modifications.

\refstepcounter{section}
\setcounter{subsection}{0}
\section*{Appendix \thesection\quad Parameter Space Determination} \label{ap:parameter_space}
This Appendix explains how we calibrate the education-resource index and how we determine the admissible parameter region used in estimation.
Throughout, it is useful to distinguish between two objects.
The first is a \textit{coarse candidate set}, used for preliminary grid search and numerical screening.
The second is the \textit{local optimization region}, which is constructed around promising grid points and is used for the continuous optimization step.
Accordingly, the final estimate need not coincide with one of the coarse grid values and may lie slightly outside the original coarse candidate set.

\subsection{The Proportion Parameter of Education Resource} \label{ap:subsec:alpha}

\begin{figure}[tp]
    \begin{minipage}[b]{\textwidth}
        \centering
        \includegraphics[width=\linewidth]{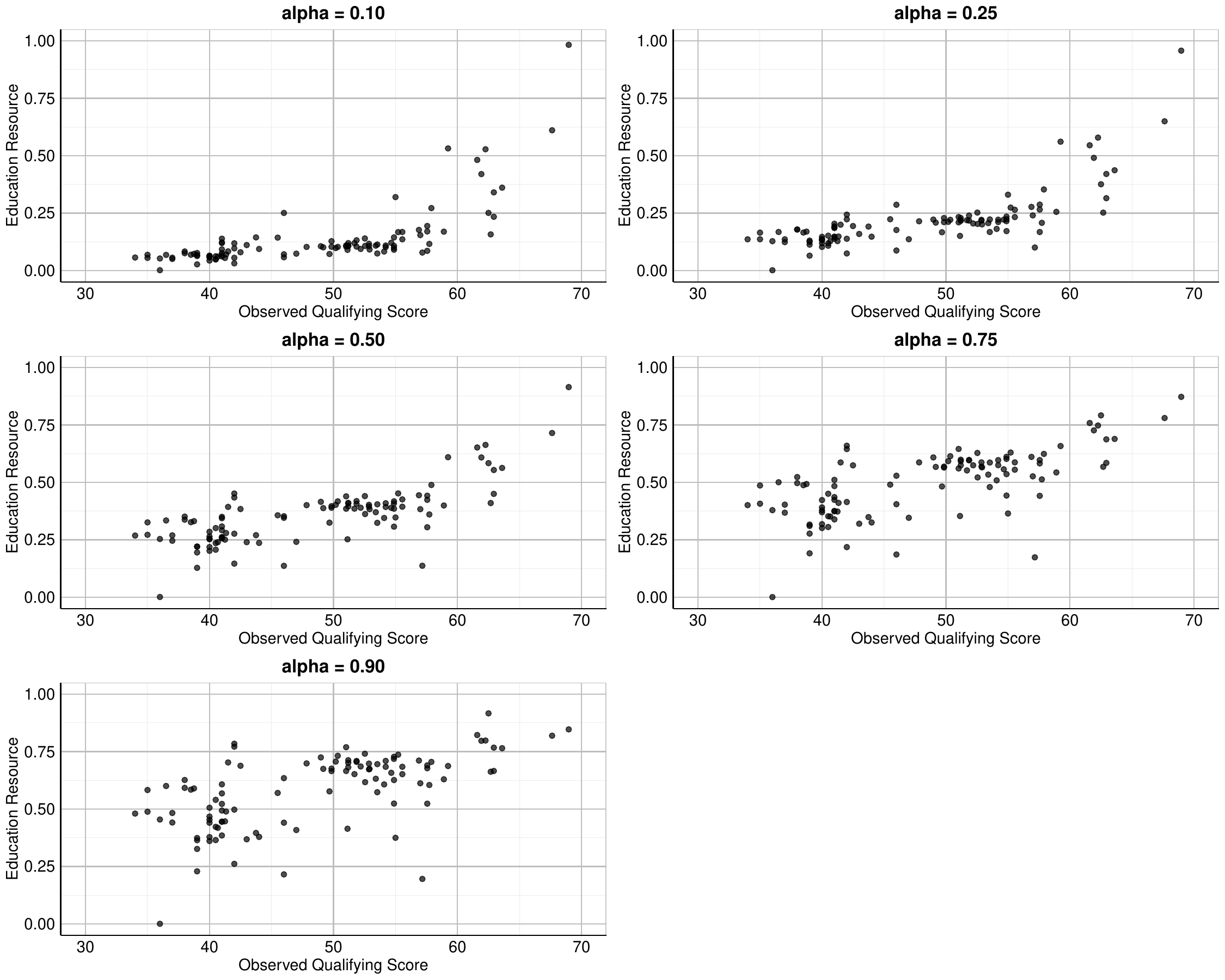}
        \captionsetup{width=.95\textwidth, font=footnotesize}
        \captionof{figure}{
        Education resource when $\alpha = 0.10, 0.25, 0.50, 0.75, \ \text{and} \ 0.90$.
        }
        \label{fig:er_calibration}
    \end{minipage}
\end{figure}

From \eqref{eq:er}, a higher value of $\alpha$ places a larger weight on the student-teacher ratio in the construction of the education-resource index.
Figure \ref{fig:er_calibration} illustrates the education resource when the proportion parameter $\alpha \in \{ 0.10, 0.25, 0.50, 0.75, 0.90 \}$.
For all candidate values of $\alpha$, the resulting education-resource index is positively correlated with the observed qualifying score, which is a basic requirement for using it as an exogenous ranking anchor in the model.
We first exclude $\alpha = 0.75$ and $0.90$, because the 1st and 2nd colleges in observed qualifying score do not have the same hierarchical ranking in terms of the education resource.
In the indirect utility equation \eqref{eq:indirect_utility}, education resource is the only exogenous component that anchors the relative ranking of colleges.
Therefore, if the education-resource ranking departs too much from the observed hierarchy of qualifying scores, the model is more likely to generate derived cutoffs that are difficult to reconcile with the data.

In the remaining candidates, $\alpha = 0.10$ is preferred because it generates the least unnecessary vertical dispersion in the education-resource index, especially among colleges with relatively low observed qualifying scores ($e_n^{\obs} \in (35,58)$).
When colleges have similar observed qualifying scores, imposing excessive exogenous dispersion in education resource would force the model to differentiate them more strongly than the data suggest.
Using $\alpha = 0.10$ therefore provides a conservative calibration that preserves the exogenous ordering of colleges while avoiding unnecessary amplification of cross-college heterogeneity.

\subsection{Parameter Space Determination} \label{ap:subsec:para_space}

In this subsection, we elucidate how we determine the parameter space $\boldsymbol{\Theta}$.
In determining the parameter space $\boldsymbol{\Theta}$, we use two criteria.
The first is economic plausibility: parameter values should imply that mobility costs, effort costs, and educational gains remain within ranges consistent with institutional features of Japan’s higher education market.
The second is numerical admissibility: candidate values should generate equilibria that are computationally stable and economically meaningful under the iteration algorithm in \eqref{eq:update}.
These two criteria play different roles across parameters.
For $\theta_1$ and $\theta_4$, the main concern is economic magnitude.
For $\theta_2$ and $\theta_3$, the main concern is their joint effect on the local stability of the equilibrium computation.
A higher qualifying score raises effort costs and thereby reduces demand, whereas it also raises the peer-effect component of utility and thereby increases demand.
The former force stabilizes the updating process, while the latter can amplify changes in demand.
Thus, the relative magnitudes of $\theta_2$ and $\theta_3$ affect whether the updating rule in \eqref{eq:update} converges to a stable equilibrium.
Accordingly, we determine their admissible range using a coarse grid search over $(\theta_2,\theta_3)$, retaining combinations that yield numerically stable and economically plausible equilibrium outcomes.
More specifically, our estimation proceeds in two stages.
First, we conduct a coarse grid search over a finite set of candidate values for each parameter and discard combinations that produce clearly implausible or numerically unstable outcomes.
Second, for each value of the scale parameter $\lambda$, we take the best grid point as an initial value and run a local continuous optimization over the box $(0.9\theta, \ 1.1\theta)$ around that grid point.
For this reason, the final estimate may differ from the coarse grid values and can lie slightly outside the original candidate set used in the preliminary grid search.
The role of the coarse candidate set is therefore to define a numerically and economically sensible search region, rather than to impose a hard bound on the final estimate.

\subsubsection{The moving cost parameter} \label{ap:subsubsec:mv}

We set $\theta_1 \in [0.5, 1.3]$.
The parameter $\theta_1$ governs the strength of generalized moving costs, which include not only pecuniary relocation and transportation costs but also nonpecuniary burdens associated with living away from home.
We choose a relatively wide interval so that inter-regional mobility is neither implausibly frictionless nor unrealistically suppressed.
In practice, values in this range imply mobility burdens that are large enough to matter for college choice, but not so large as to shut down inter-prefectural enrollment almost completely.
Our final estimate of \(\theta_1\) is well inside this interval rather than near its boundary, suggesting that the chosen range is not binding.
Table \ref{tab:mv} shows the moving cost to Tokyo from Aomori, Niigata, Kanagawa, Kyoto, and Hiroshima under different values of $\theta_1$.
These values should be interpreted as generalized annual mobility burdens in the model, rather than literal transportation expenses.
They are reported only to illustrate the economic magnitude implied by different candidate values of $\theta_1$.

\begin{table}[t!]
\centering
\caption{Moving Cost to Tokyo (JPY 1 million)}
\label{tab:mv}
\begin{tabular}{lrrrrr}
  \toprule
 & Aomori & Niigata & Kanagawa & Kyoto & Hiroshima \\
  \midrule
  $\theta_1 = 0.5$ & 2.89 & 1.26 & 0.14 & 1.83 & 3.38 \\
  $\theta_1 = 0.7$ & 4.04 & 1.77 & 0.19 & 2.56 & 4.73 \\
  $\theta_1 = 0.9$ & 5.20 & 2.27 & 0.24 & 3.29 & 6.08 \\
  $\theta_1 = 1.1$ & 6.35 & 2.78 & 0.30 & 4.02 & 7.43 \\
  $\theta_1 = 1.3$ & 7.50 & 3.29 & 0.35 & 4.75 & 8.78 \\
   \bottomrule
\end{tabular}
\begin{flushleft}
        Note:
        Distance from Tokyo: Aomori 577.3km, Niigata 252.7km, Kanagawa 27.2km, Kyoto 365.3km, Hiroshima 675.1km.
        Average disposable income: JPY 5.20 million.
    \end{flushleft}
\end{table}

\subsubsection{The effort cost parameter} \label{ap:subsubsec:ei}

\begin{figure}[tp]
    \begin{minipage}[b]{\textwidth}
        \centering
        \includegraphics[width=\linewidth]{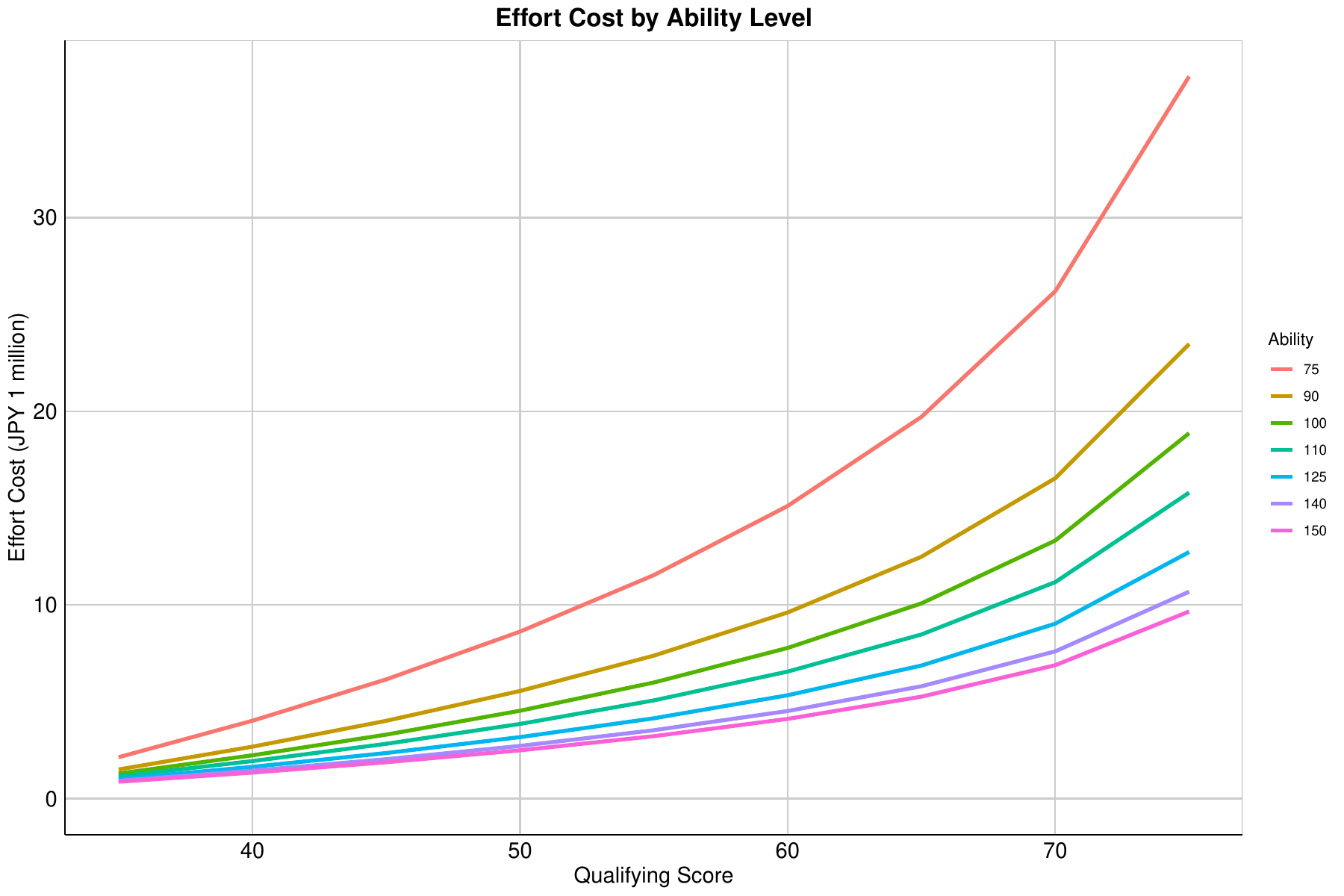}
        \captionsetup{width=.95\textwidth, font=footnotesize}
        \captionof{figure}{
        Effort cost for $a = 75, 90, 100, 110, 125, 140, \text{and } 150$ when $\theta_2 = 400$.
        }
        \label{fig:education_investment}
    \end{minipage}
\end{figure}

Figure \ref{fig:education_investment} illustrates the implied effort cost for students with different ability levels when $\theta_2=400$.
The parameter $\theta_2$ governs how rapidly the effort cost increases as students with lower ability attempt to reach higher qualifying scores.
If $\theta_2$ is too small, high qualifying scores become too easily attainable and the model tends to generate unrealistically high cutoffs across colleges.
If $\theta_2$ is too large, the implied effort costs become excessively high and the model produces implausibly weak upward mobility through study effort.
For the numerical search, we therefore restrict \(\theta_2\) to a preliminary interval $[350,450]$, which keeps the implied effort-cost schedule in a plausible range and avoids clearly unstable or unrealistic equilibrium outcomes.
Since the final estimate is obtained from the local optimization step rather than from the coarse grid itself, it may lie somewhat outside this preliminary interval.
Accordingly, the interval $[350, \ 450]$ should be interpreted as a pragmatic search range used for initialization and numerical screening, not as a hard admissible bound on the final estimate.

\subsubsection{The enrollment gain parameters} \label{ap:subsubsec:eg}

The effort-cost parameter $\theta_2$ and the peer-effect parameter $\theta_3$ jointly affect the local stability of the equilibrium iteration.
A larger $\theta_2$ strengthens the discouraging effect of higher qualifying scores through study costs, whereas a larger $\theta_3$ strengthens the positive feedback from peer quality.
If $\theta_3$ is too large relative to \(\theta_2\), the positive feedback from peer effects can dominate the negative demand response induced by higher cutoffs, making the iteration less stable numerically.
For this reason, we do not choose the admissible range of $\theta_3$ in isolation.
Instead, we use a coarse grid over $(\theta_2,\theta_3)$ and retain combinations that generate numerically stable and economically plausible equilibrium outcomes.
Within that admissible region, we set $\theta_3 \in [0.01,0.10]$ as a conservative range for the estimation.
In other words, the coarse range for $\theta_3$ is not intended to pin down the final estimate mechanically.
Its purpose is to exclude regions of the parameter space in which peer effects are so strong relative to effort costs that the equilibrium iteration becomes numerically unstable or produces implausible college hierarchies.
The peer gain ($\theta_3 \log (e_n)$) is presented in Table \ref{tab:peer_gain}.

\begin{table}[t!]
\centering
\caption{Peer Gain of Enrollment (JPY 1 million)}
\label{tab:peer_gain}
\begin{tabular}{lrrrrr}
  \toprule
 & $e_n = 35$ & $45$ & $55$ & $65$ & $75$ \\
  \midrule
$\theta_3 = 0.01$ & 0.04 & 0.04 & 0.04 & 0.04 & 0.04 \\
  $0.05$ & 0.18 & 0.19 & 0.20 & 0.21 & 0.22 \\
  $0.10$ & 0.36 & 0.38 & 0.40 & 0.42 & 0.43 \\
   \bottomrule
\end{tabular}
\end{table}

Table \ref{tab:edu_gain} shows the education gain of enrollment ($\theta_4 \log (1 + R_n)$).
In the model, $\theta_4$ serves as a classifier to stratify different level of colleges.
For example, when $\theta_4 = 20$, the return of enrolling colleges with $R_n = 0.10$ is JPY 1.91 million, and the corresponding value for colleges with $R_n = 0.90$ is JPY 12.84 million.
Meanwhile, it is also an amplifier of the derived equilibrium qualifying score.
A larger value of $\theta_4$ leads to higher enrollment gain and hence higher qualifying score of the derived equilibrium.
Thus, both excessively high and excessively low values of $\theta_4$ will create a mismatch between the derived and observed qualifying scores. 
In the coarse grid search, we use $\theta_4 \in [15, 25]$ as a preliminary candidate range.
This range is sufficiently wide to allow substantial heterogeneity in educational gains across colleges, while avoiding values that make the derived cutoff hierarchy obviously too compressed or too dispersed relative to the data.
Because the second-stage optimization is carried out over a local neighborhood around the best-performing grid point, the final estimate may lie outside this coarse interval.
Indeed, the estimated value $\hat{\theta}_4 = 27.5$ lies on the upper boundary of the local optimization region.
This suggests that, within the neighborhood examined in the second-stage optimization, the objective function was still decreasing in $\theta_4$ near the boundary.
Accordingly, the interval $[15, 25]$ should be interpreted as a conservative coarse search range used for initialization and numerical screening, rather than as a hard admissible bound on the final estimate.

\begin{table}[t!]
\centering
\caption{Education Gain of Enrollment}
\label{tab:edu_gain}
\begin{tabular}{lrrrrrrr}
  \toprule
 & $R_n = 0.10$ & $0.15$ & $0.30$ & $0.45$ & $0.60$ & $0.75$ & $0.90$ \\
  \midrule
$\theta_4 = 15$ & 1.43 & 2.10 & 3.94 & 5.57 & 7.05 & 8.39 & 9.63 \\
  $17.5$ & 1.67 & 2.45 & 4.59 & 6.50 & 8.23 & 9.79 & 11.23 \\
  $20$ & 1.91 & 2.80 & 5.25 & 7.43 & 9.40 & 11.19 & 12.84 \\
  $22.5$ & 2.14 & 3.14 & 5.90 & 8.36 & 10.58 & 12.59 & 14.44 \\
  $25$ & 2.38 & 3.49 & 6.56 & 9.29 & 11.75 & 13.99 & 16.05 \\
   \bottomrule
\end{tabular}
\end{table}

\subsubsection{The outside option parameter} \label{ap:subsubsec:outopt}

The outside option utility $u_0$ functions as a reservoir in the model.
Specifically, it affects the students' decision-making problem of whether or not to apply to a college, especially for those at the margin.
In general, a small $u_0$ leads to too many college applicants so that the derived equilibrium qualifying scores are higher than those being observed.
In the coarse grid search, we use $u_0 \in [2.5, 4.0]$ as a preliminary candidate range.
This means that the utility of not enrolling any college is JPY 2.5 to 4.0 million.
As with the other parameters, this interval is used to initialize and discipline the numerical search, while the final estimate is obtained from the subsequent local optimization step.

\subsubsection{The scale parameter} \label{ap:subsubsec:scale}

In Subsection \ref{subsec:estimation} and Remark \ref{rm:scale}, we elucidate how we estimate $\lambda$ and why this is important.
Preliminary simulation exercises suggest that values of $\lambda$ greater than one are more consistent with the observed data than values near or below one.
These diagnostic statistics include qualifying score, college vacancy rate, and inter-regional movement flows.
This suggests that the indirect utility ($V_{in}(a)$) takes a larger share than the random utility ($\varepsilon_{in}$) in the students' decision-making, implying the validity of our proposed model.
We choose $\{1.5, 1.75, 2.0, 2.25, 2.5\}$ as candidate values of $\lambda$.
Since $\lambda$ is optimized by grid search rather than local continuous optimization, the final estimate of $\lambda$ is always one of these candidate values.

\subsubsection{Numerical Admissibility Check} \label{ap:subsubsec:simul_res}

This subsection provides an illustrative numerical diagnostic for the equilibrium algorithm under a representative candidate parameter vector.
The purpose of this exercise is not to report the final estimation result, but to show how the model behaves within the preliminary admissible region used for the coarse grid search.
In particular, we use this simulation to examine whether the updating algorithm converges, whether the implied vacancy rates are economically reasonable, and whether the resulting qualifying-score pattern is broadly consistent with the data.
For this diagnostic exercise, we use $\boldsymbol{\theta} = (0.9, 400, 0.05, 20, 3.5)$ and $\lambda = 2.0$.
The simulation converges after 1,473 iterations and takes 51.6 seconds on a MacBook Pro with an Apple M3 Pro chip.

\begin{figure}[tp]
    \begin{minipage}[b]{\textwidth}
        \centering
        \includegraphics[width=\linewidth]{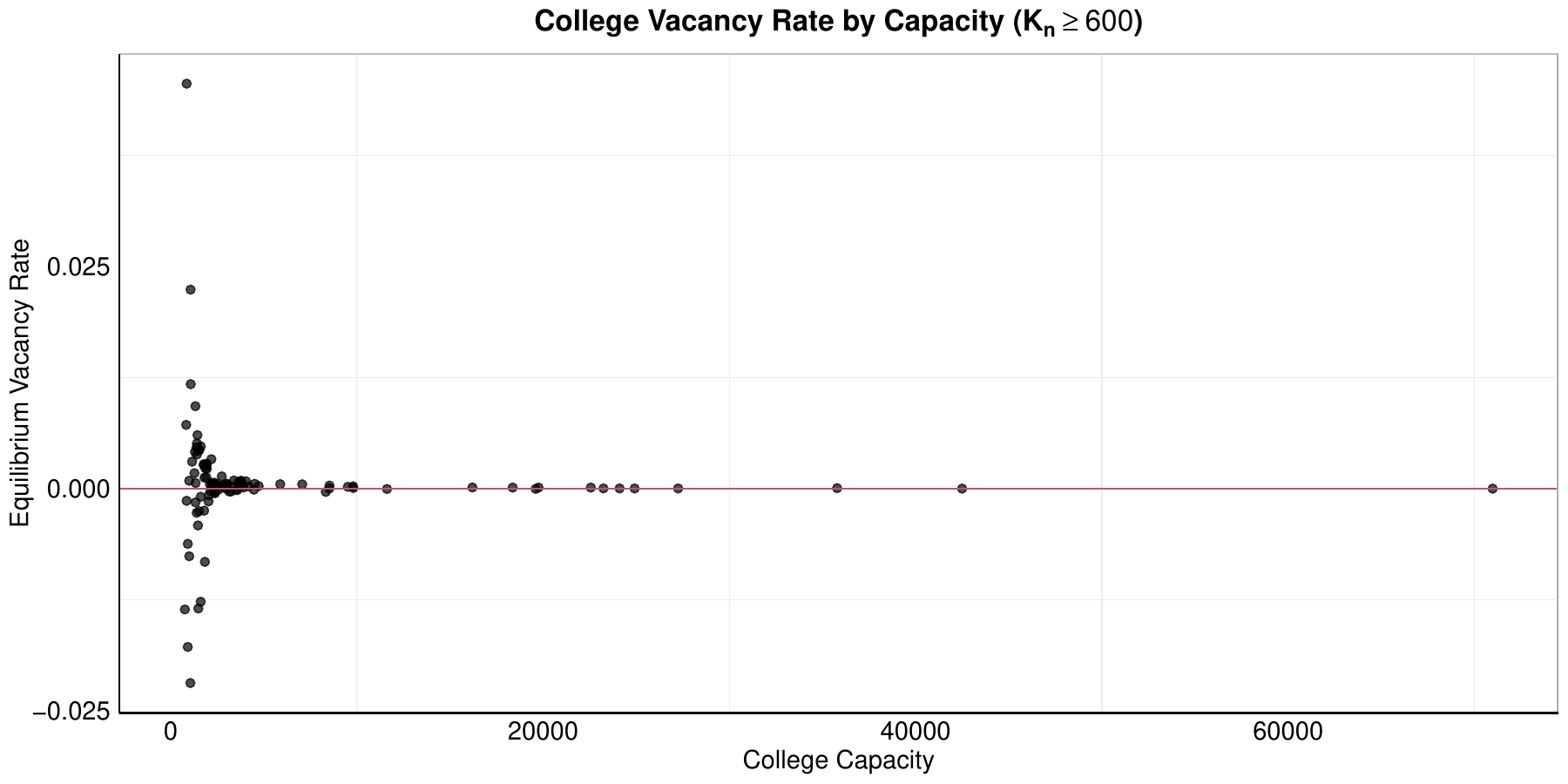}
        \captionsetup{width=.95\textwidth, font=footnotesize}
        \captionof{figure}{College vacancy rate by the simulation test.
        Negative values refer to overcrowding rate.
        }
        \label{fig:vacancy}
    \end{minipage}
\end{figure}

Figure \ref{fig:vacancy} illustrates the colleges' vacancy rate of the simulation.
From this figure, we observe that for colleges with $K_n \geq 600$, the vacancy or overcrowding rate is smaller than 5\%.
Meanwhile, the fluctuations are large in colleges with small capacities; for $K_n \geq 2,000$, the maximum vacancy rate and overcrowding rate are 0.14\% and 0.33\%, respectively.
This result suggests that the proposed model is better suited to colleges with normal or large student bodies.
For small student-body colleges, they often offer only a few majors due to their college scale.
This violates some basic assumptions of our proposed model, and researchers need to modify these assumptions to accommodate the small student-body scenario.

\begin{table}[t!]
\centering
\caption{Vacancy Rate for $K_n < 600$}
\label{tab:vacancy}
\begin{tabular}{lrrrrrr}
  \toprule
 Name & $K_n$ & $E_n$ & Vacancy Rate & $e_n$ & $e_n^{\obs}$ & $R_n$ \\
  \midrule
Akita\_Pri & 470 & 563.19 & -0.20 & 47.69 & 36.50 & 0.07 \\
Kanagawa\_Pri1 & 115 & 153.43 & -0.33 & 63.77 & 62.50 & 0.25 \\
Toyama\_Pri & 330 & 402.69 & -0.22 & 47.30 & 35.00 & 0.05 \\
Tottori\_Pri &  80 & 183.98 & -1.30 & 49.51 & 46.00 & 0.07 \\
Saga\_Pri & 540 & 571.56 & -0.06 & 54.43 & 41.00 & 0.12 \\
   \bottomrule
\end{tabular}
\end{table}

\begin{figure}[tp]
    \begin{minipage}[b]{\textwidth}
        \centering
        \includegraphics[width=0.75\linewidth]{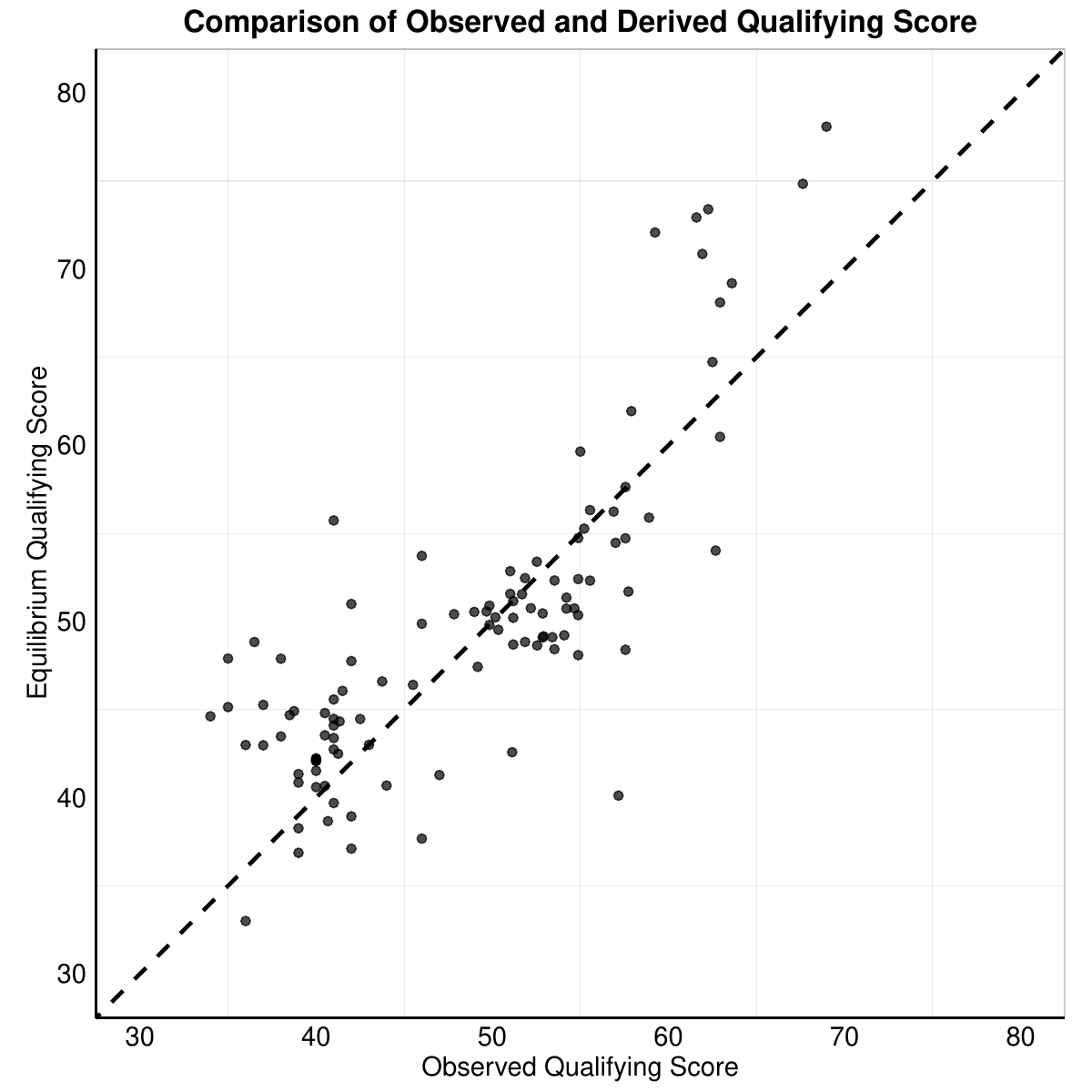}
        \captionsetup{width=.95\textwidth, font=footnotesize}
        \captionof{figure}{
        Comparison of the qualifying score between the model prediction (vertical axis) and the observed data (horizontal axis) based on the simulation in \ref{ap:subsubsec:simul_res}.
        The dashed line is $45^\degree$ line.
        Correlation coefficient: 0.805 for all colleges, 0.855 for public colleges, and 0.446 for private colleges.
        }
        \label{fig:simul_qs}
    \end{minipage}
\end{figure}

Table \ref{tab:vacancy} depicts the vacancy rates for colleges with $K_n < 600$.
From this table, we observe that all these colleges are private, all of these experience overcrowding, and $e_n > e_n^{\obs}$ for all these colleges.
Since we allow a relaxed numerical criterion for colleges with $K_n < 600$ in \ref{step:equi:check} of \ref{alg:inner}, substantial overcrowding can arise for some small colleges in the simulation; the most extreme case is Tottori\_Pri, with an overcrowding rate of 130\%.
In reality, this college only provides a nursing course, and hence it can only attract students pursuing nursing degrees.
Hence, the reason it appears as an outlier in the model is that it does not satisfy the basic assumption of students' college preference, that students choose colleges based on their qualifying scores and education resource, disregarding their capacity (and hence available majors).

Figure \ref{fig:simul_qs} illustrates the comparison of the derived and observed qualifying score under the simulation.
This figure shows desirable properties resembling its counterpart Figure \ref{fig:qs} under optimal parameters.
The value of the objective function \eqref{eq:objective} is 667,066.

\subsection{Counterfactual Entrance Rate} \label{ap:subsec:entrance_rate}

We list the college entrance rate by simulation and the difference of college entrance rate between simulation and counterfactual.

\begin{table}[p!]
\centering
\caption{College Entrance Rates under Benchmark and Counterfactual}
\label{tab:entrance_rate}
\begin{tabular}{lrrr}
\toprule
Name & Benchmark & Counterfactual & Diff \\
\midrule
Hokkaido & 42.4 & 43.1 & 0.6 \\
Aomori & 28.3 & 47.1 & 18.7 \\
Iwate & 27.2 & 51.4 & 24.2 \\
Miyagi & 49.8 & 55.7 & 5.9 \\
Akita & 27.6 & 50.4 & 22.8 \\
Yamagata & 44.8 & 56.2 & 11.4 \\
Fukushima & 35.6 & 51.0 & 15.4 \\
Ibaraki & 55.9 & 58.8 & 2.9 \\
Tochigi & 52.6 & 56.0 & 3.4 \\
Gunma & 51.8 & 54.4 & 2.5 \\
Saitama & 62.6 & 58.3 & -4.3 \\
Chiba & 62.4 & 59.4 & -3.0 \\
Tokyo & 71.7 & 66.3 & -5.4 \\
Kanagawa & 66.3 & 61.9 & -4.5 \\
Niigata & 42.3 & 56.2 & 13.9 \\
Toyama & 52.9 & 61.5 & 8.6 \\
\bottomrule
\end{tabular}

\begin{flushleft}
    Note:
    Column `Benchmark' refers to the simulated college entrance rate under the baseline equilibrium.
    Column `Counterfactual' refers to the simulated entrance rate under the no-moving-cost counterfactual.
    Column `Diff' is calculated as Counterfactual minus Benchmark and is reported in percentage points.
\end{flushleft}
\end{table}

\clearpage

\begin{table}[p!]
\ContinuedFloat
\centering
\caption[]{College Entrance Rates under Benchmark and Counterfactual (continued)}
\begin{tabular}{lrrr}
\toprule
Name & Benchmark & Counterfactual & Diff \\
\midrule
Ishikawa & 53.6 & 59.4 & 5.8 \\
Fukui & 57.6 & 61.0 & 3.4 \\
Yamanashi & 48.5 & 51.9 & 3.4 \\
Nagano & 45.7 & 55.7 & 10.0 \\
Gifu & 59.4 & 58.6 & -0.8 \\
Shizuoka & 54.5 & 58.3 & 3.8 \\
Aichi & 62.8 & 60.6 & -2.2 \\
Mie & 61.3 & 58.8 & -2.5 \\
Shiga & 69.7 & 60.0 & -9.8 \\
Kyoto & 66.5 & 56.7 & -9.7 \\
Osaka & 57.9 & 50.0 & -7.9 \\
Hyogo & 63.7 & 56.3 & -7.3 \\
Nara & 59.3 & 53.0 & -6.3 \\
Wakayama & 42.1 & 47.0 & 4.9 \\
Tottori & 42.3 & 54.7 & 12.3 \\
Shimane & 42.5 & 58.6 & 16.2 \\
\bottomrule
\end{tabular}

\begin{flushleft}
    Note:
    Column `Benchmark' refers to the simulated college entrance rate under the baseline equilibrium.
    Column `Counterfactual' refers to the simulated entrance rate under the no-moving-cost counterfactual.
    Column `Diff' is calculated as Counterfactual minus Benchmark and is reported in percentage points.
\end{flushleft}
\end{table}

\clearpage

\begin{table}[p!]
\ContinuedFloat
\centering
\caption[]{College Entrance Rates under Benchmark and Counterfactual (continued)}
\begin{tabular}{lrrr}
\toprule
Name & Benchmark & Counterfactual & Diff \\
\midrule
Okayama & 53.0 & 56.4 & 3.4 \\
Hiroshima & 55.2 & 55.1 & -0.1 \\
Yamaguchi & 40.6 & 50.9 & 10.3 \\
Tokushima & 40.7 & 49.8 & 9.1 \\
Kagawa & 45.2 & 52.9 & 7.8 \\
Ehime & 34.1 & 47.3 & 13.3 \\
Kochi & 26.9 & 45.7 & 18.8 \\
Fukuoka & 53.9 & 49.4 & -4.5 \\
Saga & 53.3 & 53.4 & 0.1 \\
Nagasaki & 30.7 & 45.8 & 15.1 \\
Kumamoto & 38.3 & 47.7 & 9.4 \\
Oita & 32.8 & 45.5 & 12.7 \\
Miyazaki & 23.4 & 43.1 & 19.7 \\
Kagoshima & 22.0 & 39.7 & 17.7 \\
Okinawa & 16.2 & 33.5 & 17.3 \\
\bottomrule
\end{tabular}

\begin{flushleft}
    Note:
    Column `Benchmark' refers to the simulated college entrance rate under the baseline equilibrium.
    Column `Counterfactual' refers to the simulated entrance rate under the no-moving-cost counterfactual.
    Column `Diff' is calculated as Counterfactual minus Benchmark and is reported in percentage points.
\end{flushleft}
\end{table}

\end{document}